\documentstyle[aps,preprint,epsfig]{revtex}
\tighten
\begin{document}  
\preprint{FIU-2328-01}
\title{\bf Selected Topics in  High Energy Semi-Exclusive Electro-Nuclear Reactions}
\author{Misak M. Sargsian}
\address{Department of Physics \\
Florida International University, Miami, FL 33199}
\maketitle
{\centerline{\today{}}
\begin{abstract}
We review the present status of the theory of high energy 
reactions with  semi-exclusive  nucleon electro-production  from 
nuclear targets. We  demonstrate how the increase of transferred energies 
in these reactions opens a complete  new window in studying the microscopic 
nuclear structure at small distances. 
The simplifications in theoretical descriptions associated
with the increase of the energies are discussed. The theoretical 
framework for calculation of high energy nuclear reactions based on 
the effective  Feynman diagram rules is described in  details. 
The result of this approach is the generalized eikonal approximation (GEA),
which is reduced to Glauber approximation when nucleon recoil is neglected.
The method of GEA is demonstrated in the calculation of high energy 
electro-disintegration of the deuteron and $A=3$ targets. Subsequently 
we generalize the obtained formulae for $A>3$ nuclei. The relation of 
GEA to the Glauber theory is analyzed. Then based on the GEA framework 
we discuss some of the phenomena which can be studied in exclusive reactions, 
these are: nuclear transparency  and short-range correlations in nuclei. 
We illustrate how light-cone dynamics of high-energy scattering
emerge naturally in high energy electro-nuclear reactions.
\end{abstract}

\section{Introduction}

Presently, our knowledge of the microscopic structure of nuclei is
limited mainly to the dynamics of single-nucleon states characterized 
by the momenta $\le 200-250~MeV/c$ and excitation energies  significantly 
smaller then the nucleon mass ($\le 100~ MeV$). The dynamics of the nuclear 
structure at very short distances in which nucleons may be  significantly 
overlapped is practically unexplored. Naive estimations indicate that such 
a configurations may provide instantaneous hadronic densities up to four 
times larger than average nuclear densities, comparable to those that may 
occur in a neutron star. However to provide a direct access to  these 
configurations the high momentum and energy should be transfered to the 
nucleus to resolve shortest possible space-time distances of the nuclear 
structure.
Providing a large values of missing momentum and energy in these processes
and using higher resolving power of the high momentum and energy transfer 
reactions one will be able to probe 
directly the short-range nucleonic correlations, which are believed to be the 
main source of high-momentum component of the  nuclear wave functions.
The fundamental  question that these studies can answer is at which 
time intervals and distances the quark and gluons become the relevant 
degrees of freedom for multinucleon configurations at small distances - 
short range correlations.

The electro-nuclear reactions are best suited for high momentum and energy 
transfer processes since the electron coupling with the virtual photon can 
be precisely calculated in QED. On the other hand the well known fact from 
particle physics that with increasing energies theoretical description of 
photon-hadron interactions become more simple and reliable 
(see e.g.  Ref.\cite{Feynman,CW}) would be relevant also for electro-nuclear 
reactions. 

In addition to the increase of the transfered momenta and energies, 
the increase of the degrees of exclusiveness of electro-nuclear reactions, 
when more products are registered in the final state of the reaction allows 
us to attain deeper  understanding of the dynamics of the reaction as well as 
to gain more information about the microscopic structure of the 
short-range nucleon correlations.

The combination of these two factors: {\em high energy and momentum transfer}
and {\em (semi) exclusiveness} of  the reaction makes electro-nuclear reactions a 
unique laboratory for studying nuclear quantum chromodynamics.
The additional boost in popularity of high-energy semi-exclusive reactions 
was given by the prediction of the unique phenomenon of 
Color Transparency (CT), 
in which the dominance of small  sized quark-gluon degrees of freedom in high 
momentum transfer $eN$ scattering is manifested by the disappearance of the 
absorption of the nucleon due to the propagation in the nuclear medium.

The last decade saw a tremendous growth of the numbers of proposed and 
performed experiments dedicated to semi-exclusive nuclear reactions with 
large values  of momentum transfer ($\ge 1~GeV/c$) (see e.g. 
\cite{PACS,Bochna,Bulten,NE18,O'Neill,Abbot,Ent,Liyanage,Airapetian,Steenhoven,Ackerstaff}). 
The experimental studies of high momentum transfer semi-exclusive reactions 
are an important part of the scientific programs of 
the high energy, high intensity electron facilities at Jefferson Lab 
(see e.g. Ref.\cite{wp}) and HERMES (see e.g. Ref.\cite{DR}). 
Therefore the development of an adequate theoretical framework to 
describe these reactions is becoming a pressing problem. 

The major issue facing the theoretical description of semi-exclusive 
reactions is that  when final state of the reaction  in addition to 
scattered electron consists of at least one hadron the strong reinteraction 
of the produced hadrons in the nuclear environment becomes the dominant feature 
of these reactions.

One of the issues in describing these reinteractions is that  the theoretical 
methods which were successful in  medium-energy  nuclear physics should be 
upgraded in order to describe hadronic reinteractions in the processes in 
which energies transferred to a nuclear target are large 
($\gtrsim few$ $GeV$). 

At energies of produced hadronic system $E_N \le 1 GeV$ the final state 
interactions  in semi-exclusive nuclear reactions are usually evaluated 
in terms of an effective potential 
for the interaction in the residual system -
the optical model approximation, (see, for example, Ref.\cite{OV}). 
Two important features of high-energy FSI  make the extension of 
the potential formalism to high energies problematic. 
Firstly, the number of essential partial waves increases rapidly 
with the energy of the $N(A-1)$ system. Secondly, the $NN$ interaction 
which is practically elastic for $E_N \le 500 MeV$ becomes predominantly 
inelastic for $E_N > 1 GeV$. 
Hence the problem of scattering hardly can be treated 
as a many body quantum mechanical problem. Introducing
in this situation a predominantly imaginary potential to 
account for hadron production is not a  well defined mathematical concept. 
So, theoretical methods successful below 1 GeV become ineffective at the 
transferred momenta which can be reached  at  Jefferson Lab \cite{CEBAF} and 
HERMES\cite{HERMES}.

Final state reinteraction in hadron induced nuclear reactions 
at higher energies ($E_N>1~GeV$) are often described within the 
approximation of the additivity of phases,  acquired in the sequential
rescatterings of high-energy projectiles off the target nucleons 
(Glauber model \cite{Glauber}). This approximation
made it possible to describe the data on elastic $hA$ scattering 
at hadron energies  $1~GeV <E_h <10-15~GeV$ (cf. Refs. \cite{Yennie,Moniz,ABV}). 

There were a many theoretical works in the last several years 
where the same principle of Glauber rescattering has been applied to  
the description of the cross sections of $A(e,e'N)(A-1)$ reactions 
\cite{F1,M1,M2,Si,Zv,Br,R2,N1,Wt,Bi,MP,Ciofi1,Ciofi2,SJ}.
Many of these works discussed the reactions in which the cross sections 
were  integrated  over the excitation  energies  of the residual 
nuclear system  and the kinematics were restricted to small  momenta of the 
residual system, $\vec p_{A-1} \le p_F$, where $p_F\approx 250~MeV/c$ is 
the characteristic Fermi momentum of the nuclear system.

In Ref.\cite{FSZ,FMSS95} the cross section of $A(e,e'N)(A-1)$ reactions 
has been calculated for  small excitation  energies that are 
characteristic for particular shells of a target nucleus with $A\gtrsim 12-16$. 

In all these cases the Glauber approximation which considers the nucleons 
in the nuclei as a stationary scatterers was a good approximation. The  
theoretical generalization for electro-production reactions was 
rather straightforward, which included mainly the account for the 
fact that energetic hadrons originate from the point of 
virtual- photon - nucleon interaction (but not from $-\infty$ as it was 
the case in $hA$ reactions).

However the Glauber approximation  can not be applied to 
the class of  $eA$ reactions in which the main emphasize is given 
to the high momentum of the bound nucleon and high excitation energies 
of the residual nuclei. This situation is especially important in studies 
of short-range nuclear properties in which large values of missing momentum 
and excitation energy are involved. 
It is also important in studies of the transparency of the nuclear medium 
for high momentum transfer electro-production reactions.

This review focuses on the development of the theoretical framework of  
calculation of the final state interactions (FSI) at high energy and momentum 
transfer (hard) semi-exclusive $A(e,e'N)X$, $A(e,e'NN)X$ reactions, with  states 
$X$ representing  ground or excited states of the  residual nucleus. 
We will concentrate on the 
kinematical region where the bound nucleon momenta and 
excitation energies in the nuclear system are large enough that 
short-range multinucleon correlations expected to  be dominant in 
the nuclear wave function. 
We are interested particularly in the region of the transfered four momenta 
$1\le Q^2 \le 6~GeV^2$. Here the lower limit is defined from the condition 
that the knocked-out hadronic system is energetic enough that the high 
energy approaches become applicable. The upper limit is defined from 
the conditions that color coherence effects are small and the produced 
hadronic state represents the single state (e.g. nucleon) but not the 
superposition of different hadronic states (wave packet)(see e.g. Ref.\cite{FMS94}).
Additionally the upper limit of $Q^2$ restricts the energies of rescattering 
nucleons, $E\le 10-15 GeV$. Later provides the ratio of inelastic diffractive to 
elastic cross sections in the  soft hadronic interactions to be a small correction. 
As it was demonstrated by Gribov\cite{Gribovi}, the smallness of this ratio is a 
necessary condition for Glauber type approximations to be a legitimate approximation.
Thus we have unique kinematic window where the theoretical methods of high energy 
scattering may  have a simpler realization due to the fact that mainly 
diagonal (elastic) terms in $hh$ rescattering will contribute.
Within this kinematic window we will discuss the apparatus of the derivation and  
application of the developed generalized eikonal approximation (GEA) to the various 
types  of semi-exclusive high energy $eA$ reactions.

The paper is organized as follows.
In Section II the general kinematic requirements are discussed. The 
emergence of the new type of small parameter in the problem is explained. 
In Section III we elaborate the basic types of the mechanisms that 
contribute to the  semi-exclusive electro-nuclear reactions. 
The high energy properties of impulse approximation in 
$eA$ reactions is considered in Section IV. 
It is demonstrated that for the selected 
components of electromagnetic current the off-shell effects 
in the bound nucleon spinors do not contribute.
In section V we discuss high energy properties of exchange currents. 
Although it is impossible to do self-consisted quantitative 
calculations still it is possible to demonstrate  based on the general 
principles that with the increase of the transferred momenta 
and energy the exchange currents will be suppressed as compared to 
the impulse approximation.
In section VI we discuss the final state interactions in semi-exclusive 
reactions and their important features in high energy limit. 
The basic points here are the emergence of the 
approximate conservation law and the reduction theorem which 
allows us to sum potentially infinite number of rescattering diagrams 
into the final set of Feynman diagrams.
The effective Feynman rules which allow to calculate
the high energy n-fold rescattering amplitude are defined in Section VII. 
These rules  constitute the basis of GEA. 
In Section VIII we demonstrate the application of the Feynman rules to 
the reaction  of high energy electrodisintegration of the deuteron. 
The GEA calculations are compared with the conventional Glauber 
approximation as well as with the calculation based on the intermediate 
energy approach. 
The application of the electrodisintegration reaction 
for the studies of color coherence phenomena in double scattering 
regime is discussed as well.
The analysis of  high energy electrodisintegration of the 
$A=3$ targets and further elaboration of  the difference between 
GEA and conventional Glauber approximation are discussed in Section IX.
In Section X the developed formalism is related to 
the studies of short-range correlations in the nuclei.
The relation of the GEA to the  
description of scattering amplitudes on the light cone is discussed in 
Section XI.
The Section XII contains the conclusions of the review.
In the Appendix A we demonstrate why closure approximation in 
general is applicable in the light cone but not in the nucleus 
rest frame at large values of missing momenta and energies. 
The details of the derivation  of the rescattering amplitude in $d(e,e'N)N$ 
reaction is given in Appendix B.

\section{General Kinematic Requirements, Emergence of  Small Parameters}
\label{sp}

We start with consideration of the general type of the reactions, 
Figure 1,  in which  high momentum $q\equiv (q_0,{\bf q})$ is transferred 
to the nucleus. The main feature of the final state in these reactions is that 
it contains the fast hadron that carries almost the entire momentum 
of the virtual photon ${\bf p_f}\approx {\bf q}$, with $|{\bf q}|\sim several~GeV/c$. 
All other hadrons ($p_s$) 
in the final state have relatively low energy ($p_s\sim~hundreds~MeV/c$).
\begin{figure}[htb]  
\vspace{-0.4cm}  
\begin{center}  
\epsfig{angle=0,width=4.8in,file=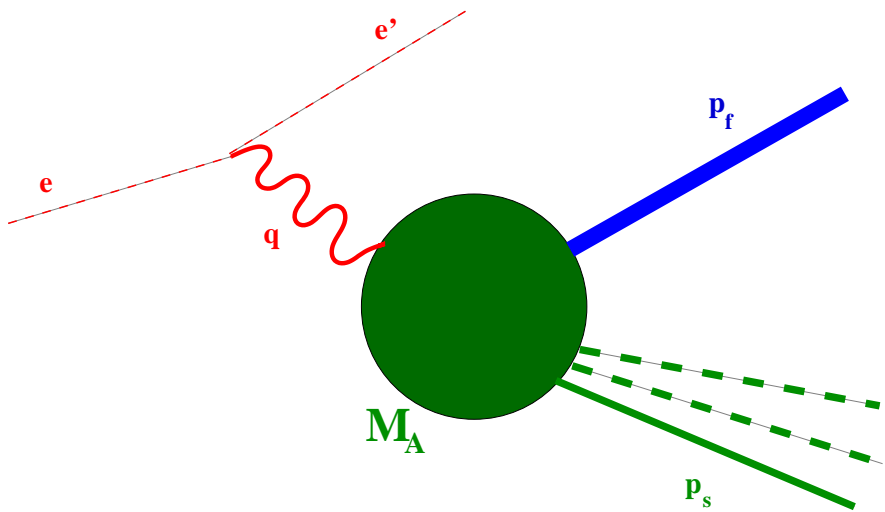}  
\end{center}  
\caption{General diagram.}  
\label{Fig.1}  
\end{figure}  
Thus the major kinematic requirement is the following:
\begin{eqnarray}
& Q^2 = {\bf q^2} - q_0^2 > 1 GeV^2,   & \ \ \  {\bf p_f} \approx {\bf q} \nonumber \\
& p_f \gg p_m, p_s \sim few \ hundreds \ MeV/c,  & \ \ \  E_f\gg E_m = E_f-q_0, 
\label{kin}
\end{eqnarray}
where ${\bf p_m} = {\bf p_f} - {\bf q}$ is the missing momentum of the 
reaction, $E_f = \sqrt{m^2+p_f^2}$ and $E_m$ characterizes the excitation of the 
residual nuclear system, $m$ is the mass of the nucleon.
Note that our definition of $E_m$ differs from the oftenly used definition of missing 
energy in which the kinetic energy of the center of mass of $A-1$ system is subtracted. 
However for our discussions this difference is not important.

Now if one constructs the $\pm$ components of four-momenta (which correspond to the 
energy and longitudinal momentum of the particle in the light-cone reference 
frame):
\begin{equation}
p_{\pm} = p_0 \pm p_z,
\end{equation}
where $z$ direction is defined by the direction of virtual photon momentum $\bf q$, then 
one  observes  that the condition of Eq.(\ref{kin}) corresponds to the 
smallness of the following combinations:
\begin{equation}
{q_-\over q_+}\approx {m x\over 2 q}\ll 1 \ \ \mbox{and} \ \ 
{p_{f-}\over p_{f+}} \approx  {m^2\over 4 p_f^2}\ll 1,
\label{small}
\end{equation}
were $x={Q^2\over 2 m q_0}$.
The availability of these {\em small parameters} is one of the important feature 
of  high energy scattering as compared to the low-intermediate energy reactions.
We will see below how these conditions  will simplify the theoretical treatment of 
semi-exclusive $eA$ reactions.

{\em Some useful rules:} Throughout the text we may use dot-product rules which 
apply to the light-cone momentum definition of $p_\pm$. 
The four momentum has two equivalent 
way of the representation $p^\mu = (p_0,p_z,p_\perp) = (p_+, p_-, p_\perp)$ and 
the scalar product of two four-vectors is defined as:
\begin{equation}
p_1\cdot p_2\equiv p_{10}p_{20} - p_{1z}p_{2z} - p_{1\perp}p_{2\perp} =
{1\over 2}p_{1+}p_{2-} + {1\over 2}p_{1-}p_{2+} - p_{1\perp}p_{2\perp}
\label{dot}
\end{equation}

\section{Some Basic Features of Semi-Exclusive Electro-Nuclear Reactions}

The first semi-exclusive electro-nuclear reactions which have been studied 
at intermediate energies confirmed the expectations that complexity of hadronic 
system significantly restricts the unambiguous treatment of underlying dynamics 
of the reaction. In general, four main processes are contributing to the  
semi-exclusive electro-nuclear reactions-Figure 2, in which at least on energetic 
nucleon is registered in the final state: 
\begin{figure}[h]  
\vspace{-0.4cm}  
\begin{center}  
\epsfig{angle=0,width=6.0in,height=1.8in,file=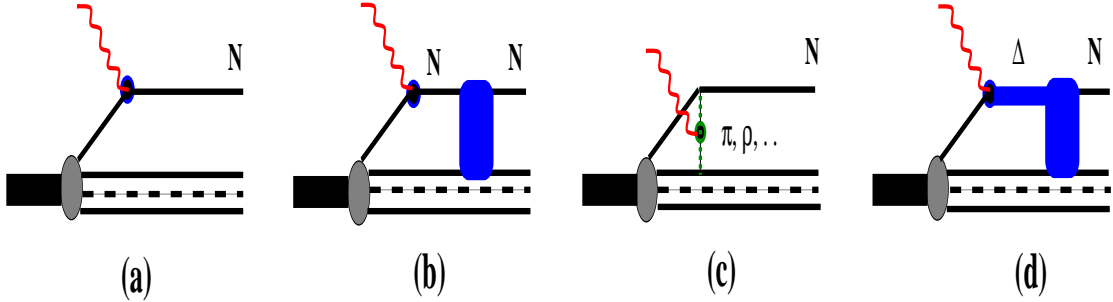}  
\end{center}  
\caption{Impulse Approximation (a), final state interaction (b), meson exchange (c), 
and $\Delta$-isobar contribution (d) diagrams.}  
\label{Fig.2}  
\end{figure}

\begin{itemize}
\item Impulse approximation (IA) amplitude (Figure 2a), 
in which the virtual photon knocks-out the bound nucleon 
which propagates to the final state without further interactions,

\item
Final state interaction (FSI) amplitude (Figure 2b), in which the knocked-out nucleon 
reinteracts  with residual hadronic system,

\item 
Amplitude with meson exchange currents (MEC) (Figure 2c), in which the  virtual photon 
interacts with the exchanged (between two-nucleon system) mesons,

\item
Isobar current contribution amplitude (IC) (Figure 2d), in which the virtual 
photon produces $\Delta$-isobar which reinteracts with residual nuclear system 
producing final hadronic state.
\end{itemize}

The study of the small distance properties of nuclei in these reactions is 
related to the exploration of IA diagram at high values of missing momenta and 
energy.
However in such kinematics at small $Q^2$ ($\ll 1~GeV^2$), practically in all cases 
the FSI, MEC and IC give dominant contributions\cite{Saclay,Mainz,NIKHEF}.

There are several reasons for the dominance of FSI, MEC and IC diagrams in the 
kinematics of large missing momentum $p_m$, missing energy $E_m$ and low $Q^2$. 
While at large $p_m$ and $E_m$ the IA amplitude  is defined by the nuclear wave 
function at  short inter-nucleon distances, the FSI, MEC and IA amplitudes  are  
defined by nuclear wave function of average configurations.

The dominance of MEC and IC amplitudes follows also 
from the kinematical considerations that at small $Q^2$ 
high missing momenta $p_m$ in semi-exclusive $A(e,e'N)X$ reactions can be observed 
only at $x<1$, i.e. in the kinematic region close to the pion threshold. It can be 
seen from Figure 3, where we calculated the $p_{mz}$ dependence on $x$ for 
quasi-elastic $e+d\rightarrow e'+p+n$ reaction that at $Q^2 < 1 GeV^2$ 
only $x<1$ is appropriate for detection of large $p_{mz}\ge 300 MeV/c$.

What  concerns to the final state interactions at small $Q^2$, they are 
dominated by  $S$ wave scattering and have broad angular distributions. Thus 
there exist no  clear criteria how to isolate or suppress FSI with respect to PWIA.

In general terms  the dominance of FSI, MEC and IA amplitudes 
means the impossibility to probe small space-time intervals 
in the nucleus using probes of larger size ($1/q \ge 1~fm$).   
\begin{figure}[th]  
\vspace{-0.4cm}  
\begin{center}  
\epsfig{angle=0,width=4.8in,height=3.6in,file=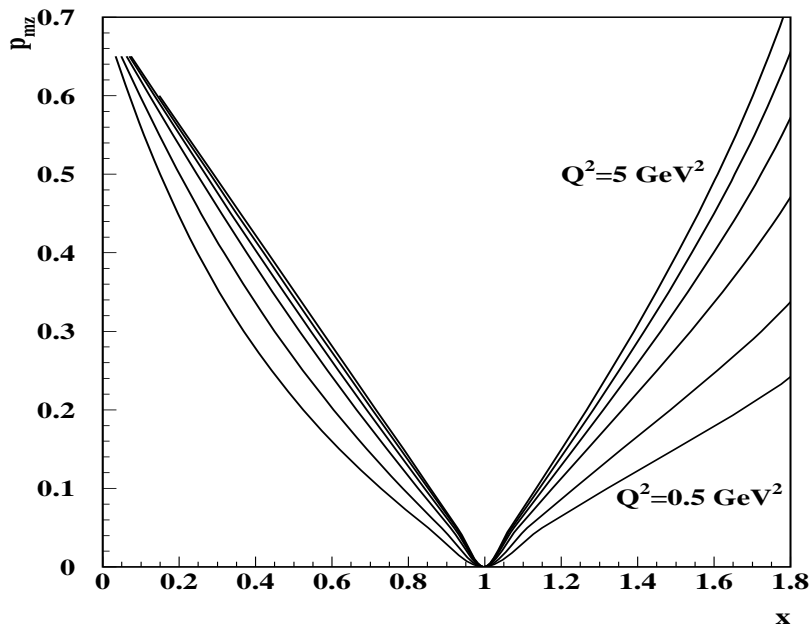}  
\end{center}  
\caption{The $|p_{mz}|$ as a function of x, for different values of $Q^2$, for 
quasielastic $d(e,e'N)N$ reaction. The lines between $Q^2=0.5$ and $Q^2=5GeV^2$ 
curves correspond to the values of $Q^2=1, 2, 3, 4~GeV^2$.}  
\label{Fig.3}  
\end{figure}

In Figure 4 we represent the typical intermediate energy measurement, 
which demonstrates how large are MEC  and IC contributions in the cross section at 
large missing momenta and low $Q^2$.
The calculations in the kinematic region of these experiments show that at large $p_m$ 
MEC and IC significantly dominate the PWIA contribution.

With the increase of energies the situation changes qualitatively. It may sound 
paradoxical but at high energy transfer, when the wavelength of the probe becomes 
much smaller than the sizes of interacting particles the situation becomes  
simpler. Using the example presented by Cheng and Wu\cite{CW}, comparison 
of the low energy with the high energy cases is analogous to the situation 
when one explores the room and in one instance  
the shapes of the objects in the room will change with the color of the light one 
uses to look on the objects  (long wavelength case)  compared to the case when the 
shapes of the objects are independent on the color of the light (short wavelength case).
 
Mathematically, the first qualitative change we encounter with increasing  
energies is the availability of small parameters discussed in Sec. \ref{sp}. 
As we will see later, the existence of these small parameters in the situation 
when interactions are strong plays an important role in the calculation of 
these reactions.

\begin{figure}[h]  
\vspace{-0.4cm}  
\begin{center}  
\epsfig{angle=0,width=3.6in,file=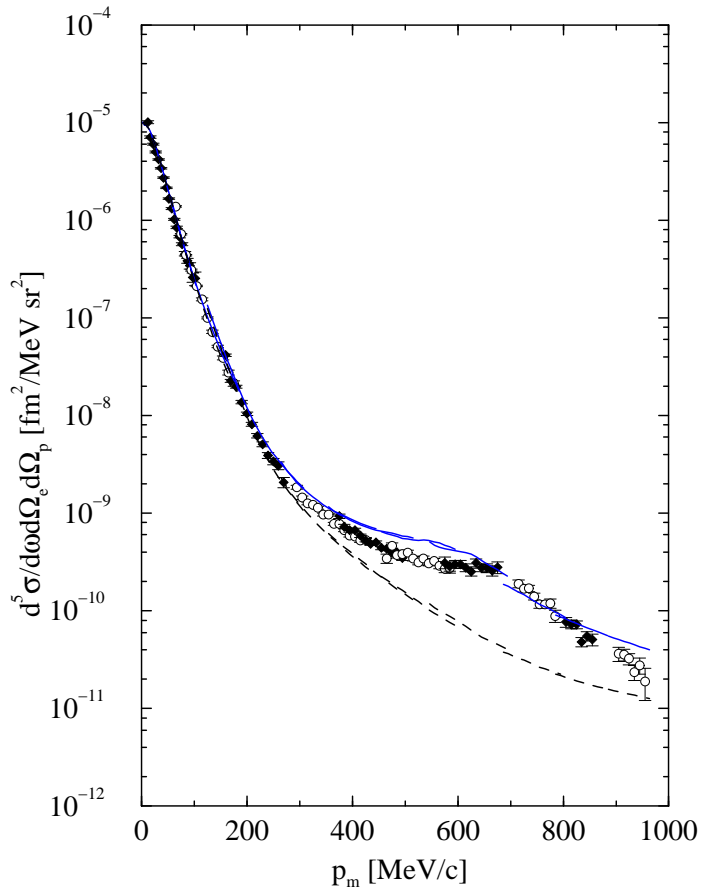}  
\end{center}  
\caption{The $p_{m}$ dependence of the differential cross sections of 
$d(e,e'p)n$ reactions at $Q^2=0.13-0.33~GeV^2$. The data are from Ref.[42]
%\cite{Mainz}. 
Solid  and dashed lines correspond to the calculation of Arenh\"ovel 
Ref.[44]
%\cite{Arenhovel}  
with and without MEC and IC contributions.
}
\label{Fig.4}  
\end{figure}

Next we consider the basic features of all of four amplitudes in Figure 2 at 
large values of $Q^2$. The impulse approximation, meson 
exchange currents and isobar contributions will be considered briefly discussing 
only the qualitative features of these amplitudes at high energies. Their 
detailed discussion is out of the scope of the present review.
The final state interaction will be considered in detail with the derivation of 
effective Feynman rules of n-fold rescattering amplitude.

\section{Impulse Approximation Diagrams}

We start with consideration of the most  simple case of electro-disintegration of the 
deuteron, Figure 5, in which the $e+d\rightarrow e^\prime+p+n$ scattering with the 
recoil nucleon momenta $p_s\sim few  \ hundred $~ MeV/c corresponds to the virtual 
photon interaction with the bound nucleon of Fermi momenta $-{\bf p_s}$. 
Choosing the momentum of bound nucleon large enough that it is no longer 
can be described as a quasifree we restrict however the upper limit of this momentum 
($\le 600~MeV/c$) so that the nucleonic degrees of freedom are still relevant 
for the bound system.
Description of the electromagnetic interaction with bound (off-shell) nucleons 
possesses many theoretical uncertainties, related to the absence of 
self-consistent theory of strong interaction that describes the binding of the 
nucleon. The origin of  the off-shell  effects in $\gamma^*N_{bound}$ scattering
amplitude is somewhat different for low and high energy domains. In the case of 
low energy transfer the nucleons represent the quasiparticles whose properties 
are modified due to the in-medium nuclear potential (see e.g. \cite{PP}). 
At high $Q^2$ the virtual photon interacts with nucleons and the phase volume of 
the process  is sufficiently large. As a result the off-shell effects in high 
energy limit are mostly related to the non-nucleon degrees of freedom. 
To demonstrate this transition as an example  we discuss the problem of 
off-shellness related to the description of the bound nucleon spinors
\footnote{Transition from quasiparticle to nucleon picture of $eA$ interaction 
could give also the natural explanation to the recently observed disappearance of 
the quenching at high values of  $Q^2$\cite{LSFSZ}}. 

\begin{figure}[h]  
\vspace{-0.4cm}  
\begin{center}  
\epsfig{angle=0,width=4.2in,file=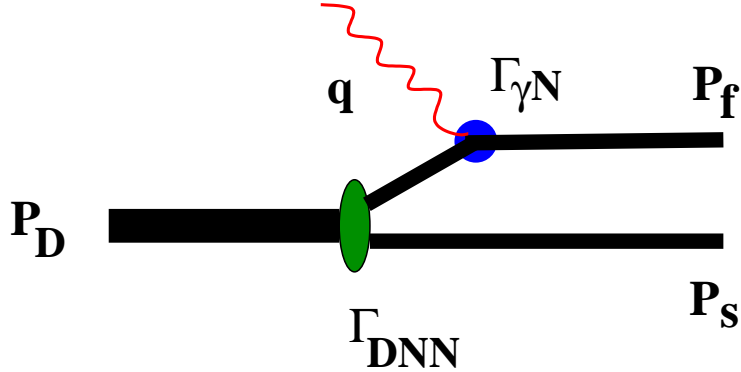}  
\end{center}  
\caption{Impulse approximation diagram for $e+d\rightarrow e'+ p + n$ scattering.}  
\label{Fig.5}  
\end{figure}
The covariant Feynman amplitude corresponding to the diagram of Figure 5 can 
be written as follows:
\begin{equation}
A^\mu_0 =  -{\bar u(p_s)\bar u(p_f) \Gamma^{\mu}_{\gamma^*N}
\cdot [\hat{p_D}-\hat{p_s}+m]\cdot \Gamma_{DNN}\over (p_D-p_s)^2-m^2 + i\epsilon}.
\label{IA}
\end{equation}
Here we used the notation $\hat{p}\equiv p_\mu\gamma^\mu$ and for simplicity  
suppressed the spin indices of the deuteron.
The $\Gamma_{DNN}$ represents the covariant $D\rightarrow NN$ transition 
vertex and $\Gamma^{\mu}_{\gamma^*N}$ is the covariant electromagnetic vertex of the 
$\gamma^* N_{bound}\rightarrow N$ transition. Note that both $\Gamma_{DNN}$ and 
$\Gamma^{\mu}_{\gamma^*N}$  are the covariant vertices and in the time ordered 
expansion they contain both impulse approximation and vacuum fluctuation diagrams 
(e.g. $\Gamma_{DNN}$ may represent $\bar N D\rightarrow N$ and 
$\Gamma^{\mu}_{\gamma^*N}$  may represent $\gamma^*\rightarrow \bar N N$). 
The lack of the self-consistent theory of strong interaction in strong QCD 
regime does not allow us to calculate both vertices from the first principles.

Furthermore we will discuss only ``$\mu=-$'' component of electromagnetic current, 
highlighting its several important features in the limit of 
${\cal O}({q_-\over q_+})$.
First, for the ``$-$'' component of the electromagnetic current the vacuum component 
of the $D\rightarrow NN$ vertex is negligible. This is true if the covariant amplitude 
is calculated in the reference frame in which the target (deuteron) has a very 
large (infinite) momentum (or in the light cone reference frame) 
(see e.g. \cite{FS81}). 
In this reference frame, for ``$-$'' component, the contribution from the 
amplitude of $\gamma^*\rightarrow N\bar N$ transition with subsequent 
$\bar N d\rightarrow N$ interaction is strongly suppressed and only the contribution 
in which $\gamma^*$ interacts with the  bound nucleon survives.
Note that this is not true for the  other components of electromagnetic currents. 
In general the complete description of the off-shell nucleon currents requires 
the negative energy state contribution with additional invariant form-factors 
as compared to the on-shell nucleon (see e.g. \cite{SG}).

The next question is, how well we can identify the $\gamma^* N_{bound}$ scattering 
in the $A_0^{-}$ amplitude of Eq.(\ref{IA}), with 
the virtual photon scattering off a free nucleon spinor. 
For this we observe that the propagator of bound nucleon in Eq.(\ref{IA}) can 
be written as follows:
\begin{eqnarray}
& & \hat p_D - \hat p_s + m = \nonumber \\ 
& & = (\hat p_D - \hat p_s + m)^{on} + 
{1\over 2}\left[(p_D-p_s)^{off}_+\gamma_- - (p_D-p_s)^{on}_+\gamma_-\right],
\label{pro1}
\end{eqnarray}
where ``off'' and ``on'' superscripts refer to the on-shell and off-shell components of 
the momentum. In derivation of Eq.(\ref{pro1}) we added and subtracted the term 
${1\over 2}(p_D-p_s)^{on}_+\gamma_-$ to construct the completely on-shell nucleon 
propagator. Using now the kinematic relations:
\begin{eqnarray}
(p_D-p_s)^{on} = {m^2+(p_D-p_s)_\perp^2\over (p_D-p_s)_-}, \nonumber \\
(p_D-p_s)^{off} = {(p_D-p_s)^2+(p_D-p_s)_\perp^2\over (p_D-p_s)_-},
\end{eqnarray}
which are based on the mass relation formula: $p_+p_- - p_\perp^2=p^\mu p_\mu$, and 
expressing the on-shell propagator through the sum of the on-shell spinors:
\begin{equation}
(\hat p_D - \hat p_s + m)^{on} = \sum\limits_{\lambda}u_\lambda(p_D-p_s)\bar 
u_\lambda (p_D-p_s),
\end{equation}
for Eq.(\ref{pro1}) one obtains:
\begin{equation}
\hat p_D - \hat p_s + m =  \sum\limits_{\lambda}u_\lambda(p_D-p_s)\bar 
u_\lambda (p_D-p_s) 
+ {(p_d-p_s)^2 - m^2\over 2(p_D-p_s)_-}\gamma_-.
\label{pro2}
\end{equation}
Inserting Eq.(\ref{pro2}) into Eq.(\ref{IA}), for ``$-$'' component of $A^\mu_0$ one 
obtains
\begin{equation}
A^{-}_0\equiv A^{on,-}_0 + A^{off,-}_0 = 
-\sum\limits_{\lambda} {\bar u(p_s)j^-(Q^2)\bar u(p_D-p_s)\Gamma_{DNN}\over 
(p_D-p_s)^2-m^2 
+ i\epsilon} - {\bar u(p_s)\bar u(p_f)\Gamma^{-}_{\gamma^*N}\cdot 
\gamma^{-}\Gamma_{DNN}\over 
2(p_D-p_s)_-}.
\end{equation}
To illustrate the suppression of  the contribution of off-shell effects in 
the spinor structure of $A^{-}_0$ we shall compare
$A^{off,-}_0$ with $A^{on,-}_0$. For this we assume the following analytic 
form of the nucleon electromagnetic vertex:
\begin{equation}
\Gamma^{\mu}_{\gamma^*N}\sim A\cdot \gamma^\mu + B\cdot 
\sigma^{\mu,\nu}q^\nu 
\label{vert}
\end{equation}
where $A$ and  $B$  are functions of  scalar combinations of 
$q^\mu$, $p_f^\mu$ and $(p_D-p_s)^\mu$.
Using this form and the identity $\gamma_-\gamma_-=0$ one can estimate that:
\begin{equation}
{A_0^{off,-}\over A_0^{on,-}}\approx {\cal O}({q_-\over q+})\ll 1.
\label{offon}
\end{equation}
Therefore for the ``$-$'' component of the IA amplitude the off shell part 
in the nucleon spinors decreases with the increase of transferred energy. 
Notice that ``$-$''component of the current in particle physics  usually called 
a ``{\em good}'' component of the current (see e.g. \cite{GMD}). It can be 
unambiguously identified in electro-nuclear reactions (for example, in the   
inclusive $A(e,e')X$ reactions the  structure function  $W_{2}\sim |A_0^{-}|^2$).

\medskip
\medskip
{\em The conclusion} of the above discussion is that in the high energy limit one can 
identify the ``good'' component of electromagnetic current which is insensitive to 
the  off-shellness of the bound nucleon spinor.  In this limit, for IA term, one obtains:
\begin{equation}
A_0^{-} \approx \psi_D(p_m)\cdot J^{-}_N(Q^2,p_f,p_m),
\label{IAap}
\end{equation}
where ${\bf p_m} = {\bf p_D}-{\bf p_s} = {\bf p_f} - {\bf q}$
and $\psi_D(p_m)$ is 
the wave function of the deuteron defined in the reference frame where the deuteron is 
fast (or in the light-cone). Thus price we pay for gaining simplicity in the 
electromagnetic current is the necessity to calculate the nuclear wave functions 
in the light cone reference frame (see e.g Refs.\cite{FS81}).

It is worth noting that in the nuclear rest frame 
one can identify the $\Gamma^{\mu}_{\gamma^*N}$
vertex with nonrelativistic wave function (see e.g.\cite{Gribov}) and electromagnetic 
current with the  positive energy nucleon spinors (see e.g. \cite{Grossi}) 
only in the limit of ${\cal O}({p_s^2\over m^2})$.  Behind this limit in 
the nuclear rest frame description,  there is no defined strategy to discriminate 
between  on- and off- shell  contributions of electromagnetic current
(for discussions see  Ref\cite{KochPol,deFor}).

It should be emphasized that above discussion on off-shellness is relevant only 
for the situations in which the nucleons are the  relevant degrees of freedom. 
At the transferred energies when nucleon substructure plays a dominant role the 
different type of off-shellness related to the medium modification of quark-gluon 
distributions emerges (e.g. EMC effects in deep-inelastic $eA$ scattering).

\section{Meson Exchange Currents and Isobar Contributions}

Next we consider the contributions of MEC and IC amplitudes in the high energy limit of 
Section \ref{sp}~(Eq.(\ref{kin})). Since both MEC and IC are two-body currents, 
it is sufficient to consider the deuteron target (Figure 6) to arrive to the general 
conclusion. 

The experimental results at low $Q^2$ $d(e,e'N)N$ data demonstrated that with 
an increase of $p_m$ the MEC and IC contributions become dominant. 
However one  expects again major qualitative 
changes with the increase of $Q^2$ ($\ge 1 GeV^2$).

\begin{figure}[h]  
\vspace{-0.4cm}  
\begin{center}  
\epsfig{angle=0,width=6.0in,file=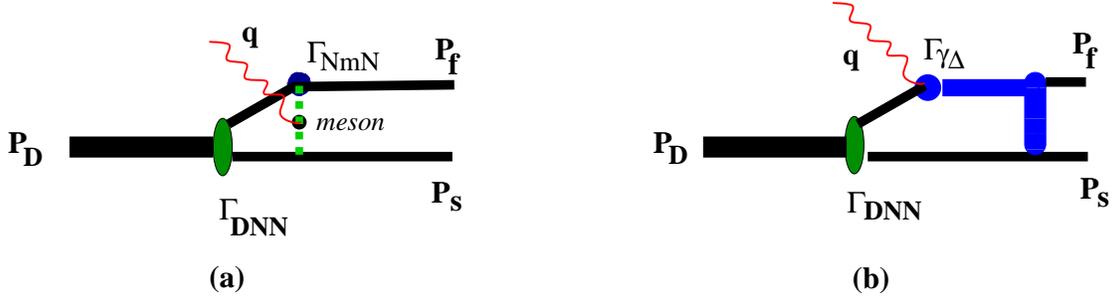}  
\end{center}  
\caption{Meson Exchange Current and $\Delta$-isobar contribution amplitudes
for $e+d\rightarrow N+N$ reaction.}  
\label{Fig.6}  
\end{figure}

The major problem we face in the estimation of MEC at high energies is that with an  
increase of $Q^2$ the virtuality of exchanged mesons in Figure 6a grows and 
$Q^2\gg m_{meson}^2$. 
(As it was noted by Feynman `` {\em pion far off its mass shell} may be a meaningless - 
or at least highly complicated idea''\cite{Feynman}).

However theoretically from very general principle, one expects that the contribution 
of meson exchange currents will decrease with an increase of 
the virtuality of exchanged photon, -$Q^2$. 
Indeed it can be shown that MEC diagrams (Figure 6a) have  an additional 
$\sim 1/Q^4$ dependence  as compared to the IA diagrams of Figure 5). 
This suppression comes from two major factors:
First, since at the considered kinematics the knocked-out nucleon is fast 
and takes almost the entire momentum of the virtual photon $q$, 
the exchanged meson propagator is proportional to $(m^2_{meson}+Q^2)^{-1}$.
Secondly, an additional $Q^2$ dependence comes from the $NN-meson$ form-factor 
$\sim (1+Q^2/\Lambda)^{-2}$. Thus overall $Q^2$ dependence of MEC amplitude 
can be estimated as follows:
\begin{eqnarray}
A_{MEC}^\mu & \sim &  \int d^3p\cdot \Psi(p){J_m^\mu(Q^2)\over 
(Q^2+m_{meson}^2)}\Gamma_{MNN}(Q^2)
\nonumber \\
& \propto & \int d^3p\cdot \Psi(p)\left({1\over (Q^2+m_{meson}^2)^2 
(1+Q^2/\Lambda^2)^2}\right)
\label{MEC}
\end{eqnarray}
where $J_m^\mu(Q^2)$ is the meson electromagnetic current proportional 
to the elastic form factor of the  meson $\sim {1\over Q^2 +m_{meson}^2}$, 
$m_{meson}\approx 0.71~GeV$ and $\Lambda\sim 0.8-1~GeV^2$
{\footnote{We assume here that different meson-nucleon vertices
have similar dependencies on the virtuality and assume the dipole 
dependence on the virtuality corresponding to neglecting the size 
of a meson as compared to the baryon size (for large $Q^2$ quark 
counting rules lead to $\Gamma_{MNN}(Q^2) \sim {1\over Q^6})$ and 
use restrictions on the Q-dependence of the 
 $\pi NN$ vertex from the measurement of the anti-quark 
distribution in nucleons, for the recent discussion
see \cite{Thomas}.}. Comparing Eq.(\ref{MEC}) with Eq.(\ref{IAap})
we can estimate the overall additional $Q^2$ dependence  as compared to  
IA contribution as: $\sim (1+Q^2/\Lambda^2)^{-2}$ which is rather 
conservative estimation (see the  footnote in this page).
Thus one expects that MEC contributions will be strongly 
suppressed as soon as $Q^2\ge (m^2_{meson}, \Lambda) \sim 1 GeV^2$.
The indication of such suppression is clearly seen in the SLAC experiment 
\cite{Frodyma}, where the ratios of structure functions, $W_1/W_2$
are measured at the deuteron threshold $x\rightarrow 2$. Since in the 
limit of  $x\rightarrow 2$ one expects the dominance of MEC\cite{Frodyma} 
(meson exchanges needed to provide the bound pn final state), 
the $Q^2$ dependence of this ratio at fixed relative energy of 
final $pn$ system will be very sensitive to the $Q^2$ dependence of 
MEC contributions. 
\begin{figure}[h]  
\vspace{-0.4cm}  
\begin{center}  
\epsfig{angle=0,width=4.2in,file=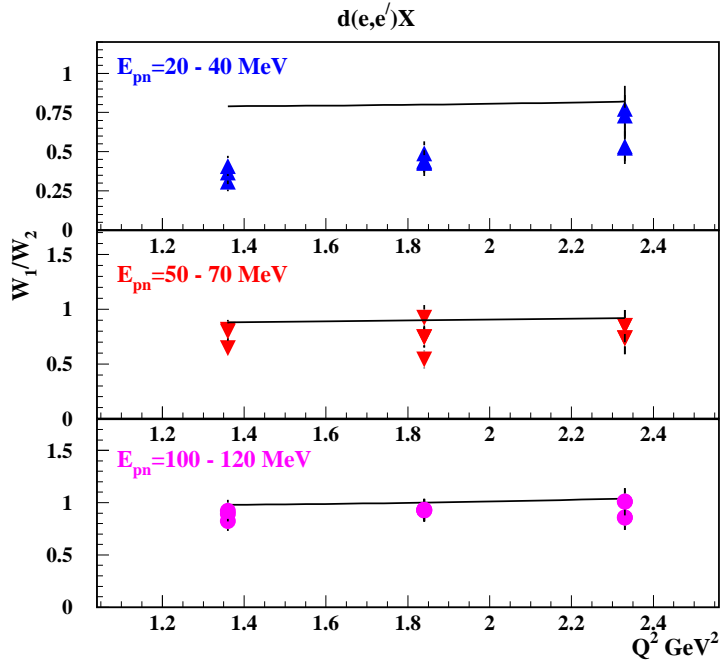}  
\end{center}  
\caption{The $Q^2$ dependence of $W_1/W_2$ from inclusive $d(e,e')X$ reaction. 
The data are from Ref.[55].
%\cite{Frodyma}.
$E_{pn}$  $(= W_{pn}-m_D-\epsilon_d^{bound}$), 
where $W^2_{pn} = W^2_{\gamma^*D} = (q^\mu + p_D^\mu)^2$, is the C.M. energy of the 
proton and neutron in the final state of the reaction. Solid lines are PWIA prediction
within light cone dynamics in collinear approach [56,101].}
%\cite{FS88,Day}
 
\label{Fig.7}  
\end{figure}

As an experimental indication of MEC suppression at high $Q^2$ in 
Figure \ref{Fig.7} we demonstrate the $Q^2$ dependence of 
$W_1/W_2$ for different values of $E_{pn}$. 
Note that $E_{pn}=0$ corresponds to the deuteron threshold where MEC should 
be especially enhanced.
The figure clearly indicates that MEC contribution decreases with increase of $Q^2$
and additionally it indicates that starting at $E_{pn}\ge 50~MeV$ MEC 
contribution becomes negligible at $Q^2\ge 1.5~GeV^2$ region. 

\medskip

For the case of  IC contribution, similar to the consideration of 
MEC,  we are interested mainly in the 
energy dependence of scattering amplitude as compared to the IA amplitude, $A_0$. 
For this it is sufficient to observe that one can estimate the scattering amplitude 
of Figure 6b as follows:
\begin{equation}
A_{IC}^\mu\sim i\int \psi_D(p_{mt}-k_t, p_{mz}-{m_\Delta^2 - m^2\over 2q}))
J_{\gamma^*N\Delta}(Q^2) A_{\Delta,N\rightarrow NN}(k_t)d^2k_t,
\label{AIC} 
\end{equation}
where $J_{\gamma^*N\Delta}(Q^2)$ and  $A_{\Delta,N\rightarrow NN}$ are 
electromagnetic $N\rightarrow \Delta$ and  hadronic $N\Delta\rightarrow NN$ 
transition amplitudes respectively. 
Comparing $A_{IC}$ with $A_0$ of Eq.(\ref{IAap}) one first observes that the 
$x < 1$ and $x > 1$ regions at  same $|p_m|$ have 
a different contributions from the intermediate 
$\Delta$ excitations. It can be seen from Eq.(\ref{AIC}) that the amplitude of 
IC contribution is proportional to:
\begin{equation}
\psi_{D}(p_{mt}-k_t, p_{mz}-{m_\Delta^2 - m^2\over 2q}), 
\end{equation}
where $p_{mt}$ and $p_{mz}$ are the transverse and longitudinal 
components of the measured missing momentum.
The Eq.(\ref{AIC}) shows that more IC contribution one has
in $x < 1$ region than $x > 1$ since in the first case  
$p_{mz}-{M_\Delta^2 - M_N^2\over 2q} < p_{mz}$, while 
for $x>1$, $|p_{mz}-{M_\Delta^2 - M_N^2\over 2q}| > |p_{mz}|$.

Therefore the access to the $x>1$ region 
at high $Q^2$ (Figure 3) allows us to separate regions where the IC contribution is 
suppressed kinematically as compared to the IA contribution.

Another possible suppression comes from the comparison of elastic 
$J_{\gamma^*N}^\mu$ and $\Delta$ transition $J_{\gamma^*N\Delta}$ currents. 
Such a suppression at high $Q^2$ is expected from the perturbative QCD arguments, 
since $\Delta$ represents helicity-flip transition which is suppressed 
in the domain of pQCD. The hypothesis on smooth matching  of nonperturbative and 
perturbative regimes of QCD suggests that the suppression of helicity flip amplitudes 
should be observed already at relatively high $Q^2\ge few ~GeV^2$. 
Indeed the experimental data on elastic and transition form factors suggest such a 
suppression\cite{Stoler}.

And finally the additional suppression of the IC contribution 
comes from the energy dependence of 
$A_{\Delta,N\rightarrow NN}$ amplitude. To estimate the energy dependence of this 
amplitude one observes that the Feynman amplitude of scattering process 
which proceeds by an exchange of particle with spin $J$ behaves as 
$A\sim s^{J}$\cite{Azimov,CW}, in which $s$ is the 
square of the center of mass energy of final NN system 
($s\approx ({2\over x}-1)Q^2+2m^2$). In the QCD description, in which 
hadrons are bound states of quarks and gluons $J$ is the Regge trajectory with 
quantum numbers permitted for a given process. Some trajectories are known 
experimentally: $J_\pi(t)\approx \alpha^\prime t$, 
$J_\rho(t) \approx 0.5 + \alpha^\prime t$, 
where $\alpha^\prime\approx 1~GeV^{-2}$ and $t$ is momentum transfer 
(see e.g.  \cite{PDBC}).

Since the transition in $A_{\Delta,N\rightarrow NN}$ is dominated by pion ($J=0$) or 
$\rho$-type 
reggeon  $(J=1/2)$ exchange  the $\Delta N\rightarrow NN$ transition amplitude will 
be suppressed at least by factor of ${1\over \sqrt{Q^2}}$ as compared to the elastic 
$NN\rightarrow NN$ amplitude. Thus IC amplitude will be suppressed additionally at 
least  by factor of ${1\over \sqrt{Q^2}}$. 
This is the upper limit in the estimation of 
IC contribution, since the data on $NN\rightarrow N\Delta$ scattering 
indicate that the $\rho$ regime  is unimportant up to  ISR energy 
range,  thus in the $Q^2$ range of our interest 
more rapid suppression  is expected with the increase of $Q^2$.

It is interesting to note that the $Q^2$ dependence of electromagnetic form 
factors and energy dependence of soft rescattering amplitudes of the other 
resonances (as $N^*$) are similar to the elastic form-factor of nucleon and  $NN$ 
soft rescattering amplitude (corresponding to the exchange of $J=1$ reggeon) 
respectively. This fact as we will discuss later has an important 
implication in the emergence of the new regime, in which the quark-gluon degrees of 
freedom of the nucleons play an important role in hard exclusive nuclear reactions.

\section{Final State Interactions}

With an increase of energies the  characteristics of soft (small angle) 
hadron-hadron  interactions in the amplitude of Figure 2b simplify in several ways.
The first major qualitative change is the emergence of practically energy 
independent  total cross section of hadron-hadron interactions at lab momenta 
$\ge 1-1.5 \ GeV/c$ (total cross sections are being constant up to momenta of 
400 $GeV/c$). 
\begin{figure}[h]  
\vspace{-0.4cm}  
\begin{center}  
\epsfig{angle=0,width=4.2in,file=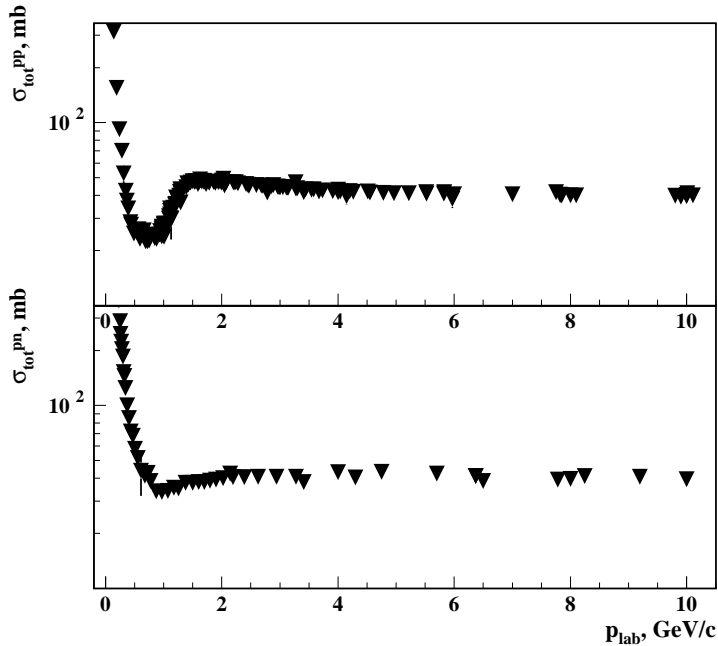}  
\end{center}  
\caption{Proton-proton and proton-neutron total cross sections as a function of incoming 
proton momenta in Lab., Ref.[60].
%\cite{PDG}.
} 
\label{Fig.8}  
\end{figure}
As Figure 8 shows both $pp$ and $pn$ total cross sections level out and become 
practically energy independent at lab  momenta greater than a $few$ GeV/c. This 
simplifies tremendously the description of the final state interactions 
since the small angle $NN$ scattering is proportional to $\sigma_{NN}^{total}$.
The additional simplification associated with the increase of energies is that 
at small angles the rescattering amplitude is predominately imaginary and conserves 
the helicity of interacting particles.

Finally, another consequence of the high energies is the onset of new 
(approximate) conservation law, which, as we will see later, has a significant 
impact in analyzing of the semi-exclusive reactions.

\subsubsection{New (Approximate) Conservation Rule}
In general, when a fast nucleon propagates through the nuclear medium it 
reinteracts with the target nucleons and as a result the information about  
preexisting momentum distribution of the bound nucleons is distorted. 
Therefore  due to FSI the reconstructed 
missing momenta does not coincide with the actual momenta of the bound nucleon in the 
nucleus (for example ${\bf p}^\prime_{m}\ne {\bf p}_m = {\bf p}_f -{\bf q}$ 
in Figure 2b).
Thus FSI can severely hinder our ability to study the nuclear interior without much 
disturbing it. However as we will see below, the increase of the energy of 
propagating nucleon allows to preserve some information about the momentum 
distribution of bound nucleon.

For this let us consider the propagation of a fast nucleon with four-momentum 
$k_1= (\epsilon_1,k_{1z}, 0)$ through the nuclear medium. We chose the $z$ axis 
in the direction of ${\bf k_1}$ such that 
${k_{1-}\over m}\approx {m\over 2k_{1z}} \ll 1$. 
After the small angle rescattering of this 
nucleon with the bound nucleon of four-momentum $p_1=(E_1,p_{1z},p_{1\perp})$, 
the energetic nucleon still attains its high momentum and leading  $z$ direction 
having now the four-momentum $k_2=(\epsilon_2,k_{2z},k_{2\perp})$ with 
${k_{2\perp}^2\over m^2_N}\ll 1$ and 
the bound nucleon four-momentum  becomes $p_2=(E_2,p_{2z},p_{2\perp})$. 
The energy momentum conservation for this scattering allows us to write for 
the ``$-$'' component:
\begin{equation}
k_{1-} + p_{1-} = k_{2-} + p_{2-}.
\label{claw}
\end{equation}
From Eq.(\ref{claw}) for the change of the ``$-$'' component of the bound 
nucleon  momentum one obtains
\begin{equation}
{\Delta p_{-}\over m} \equiv  {p_{2-}-p_{1-}\over m}\equiv \alpha_2- \alpha_1  
= {k_{1-}-k_{2-}\over m}\ll 1,
\end{equation}
where we defined $\alpha_i = {p_{i-}\over m}$, $i=1,2$ and used the fact that 
${k_{2\perp}^2\over m^2_N},{k_{1\perp}^2\over m^2_N} \ll 1$.
Therefore $\alpha_1 \approx \alpha_2$. 
The later indicates that with the increase of 
energies a new conservation law  emerges according to which the $\alpha$ of the bound 
nucleon is conserved. The uniqueness of the high energy rescattering is in the fact 
that although both energy and momentum of bound nucleon are distorted due to 
rescattering, the combination of $E-p_z$ is not. 

In Figure 9 we demonstrate the accuracy of this conservation law for momenta of a  
propagating nucleon relevant for our considerations. 
It can be seen from Figure 9 that the smaller the transferred momentum during  
rescattering, the  better is the accuracy of the conservation. It is important to 
note that in the kinematic region given by Eq.(\ref{kin}) the average transfered 
momentum in the rescattering amplitude is $<k^2_t>\approx 0.250~GeV^2/c^2$, thus 
starting from $3-4~GeV/c$ momenta of the  propagating nucleon the conservation of 
$\alpha$ ($\sim {\cal O}(1)$)  is accurate on the level of less than $5\%$.

The fact that information on the  $\alpha$ distribution of bound nucleons is 
preserved by FSI will play an important role in the  investigation of short-range 
properties of nuclei in semi-exclusive reactions. Finally it is worth mentioning that 
at small angles transferred momenta in the rescattering is practically transverse 
and its scattering amplitude can be parameterized as 
$f_{NN} \approx is\sigma_{tot}e^{-{B\over 2}p_\perp^2}$.

\begin{figure}[t]  
\vspace{-0.4cm}  
\begin{center}  
\epsfig{angle=0,width=4.2in,file=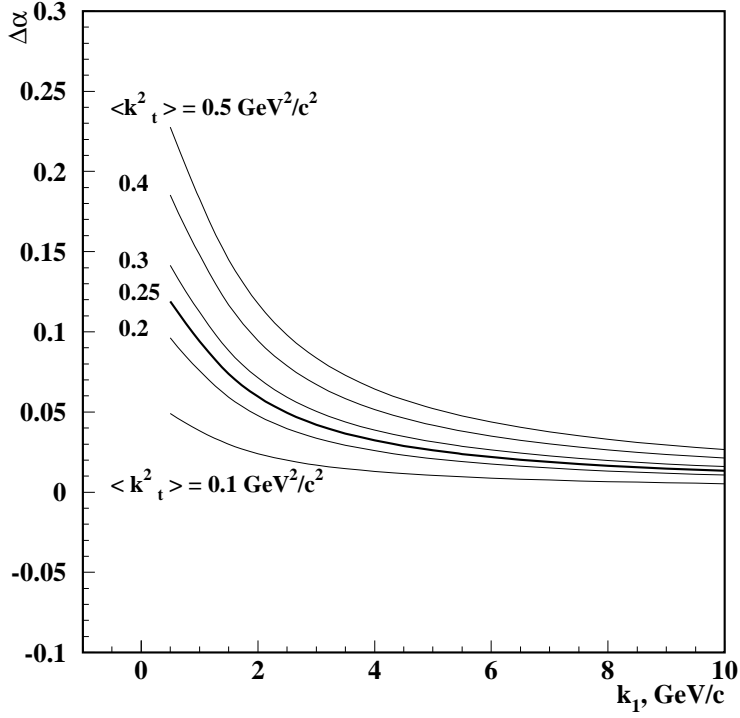}  
\end{center}  
\caption{The accuracy of the conservation of $\alpha$ as a function of the 
propagating nucleon momentum, $k_1$ at different values of average transferred 
(during the rescattering) momenta, $<k^2_t>$.} 
\label{Fig.9}  
\end{figure}

\subsubsection{Reduction Theorem}
Next we will consider the following {\bf theorem}: {\em High energy particles propagating 
in the nuclear medium can not interact with the same bound nucleon a second time after 
interacting with another bound nucleon.} In other words all those rescatterings which 
have a segments similar to Figure 10 are suppressed.

\begin{figure}[h]  
\vspace{-0.4cm}  
\begin{center}  
\epsfig{angle=0,width=4.2in,file=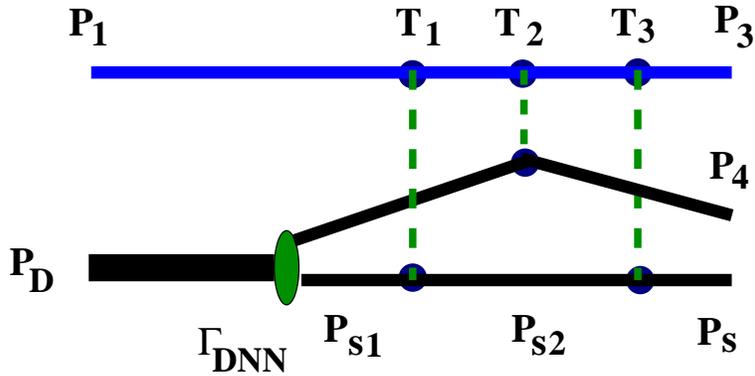}  
\end{center}  
\caption{Diagram for hadron-deuteron scattering.} 
\label{Fig.10}  
\end{figure}
We prove the above statement for  the case of the amplitude of the diagram of Figure 10. 
This prove can be easily  generalized to all other cases. 
The invariant amplitude of the diagram in Figure 10 can be presented as follows:
\begin{eqnarray}
A_{pd\rightarrow ppn} = -\int {d^4p_{s1}\over i(2\pi)^4}{d^4p_{s2}\over i(2\pi)^4}
{T_3(p_s-p_{s2})T_2(p_4-(p_D-p_{s1}))T_1(p_{s2}-p_{s1})\over 
D(p_3+p_s-p_{s2})D(p_1+p_{s1}-p_{s2})} \nonumber \\
{\Gamma_{DNN}\over D(p_{s2})D(p_{s1})D(p_D-p_{s1})},
\label{rt1}
\end{eqnarray}
where $D(p) = -(p^2-m^2+i\epsilon$), we neglect the spins since they are not relevant 
for our  discussion. 

The kinematics for the scattering corresponding to Figure 10, is such that 
$ p_{1+}\sim p_{3+} \gg m$ and $p_{1-}\sim p_{3-}\ll m$, i.e. the propagating nucleon 
in the top of the Figure 10 is very energetic. 
While $p_{4+}\sim p_{4-}\sim p_{s+}\sim p_{s-}\sim m$, i,e, recoiled nuclear 
system is non relativistic.

Next we introduce momenta:
\begin{equation}
k_1 = p_{s2}-p_{s1}, \ \ \ \ \ \  k_2 = p_s-p_{s2} \ \ \ \ \  
\mbox{and} \ \ \ \ \ \  K = k_1 + k_2.
\label{momenta}
\end{equation}
Using these definitions and the relation $d^4K = {1\over 2} d^2K_\perp dK_- dK_+$ 
for Eq.(\ref{rt1}) we obtain:
\begin{eqnarray}
& & A_{pd\rightarrow ppn} = {1\over 4}\int {d^2K_\perp d^2k_{2\perp}\over (2\pi)^4}
 {dK_- dk_{2-}\over (2\pi)^2} {dK_+ dk_{2+}\over (2\pi)^2}\nonumber \\ 
& & {T_3(k_2)T_2(p_4-(p_D-p_{s}+K))T_1(K-k_2)\over 
D(p_3+k_2)D(p_1+k_2-K)}  
%\nonumber \\ 
{\Gamma_{DNN}\over D(p_{s}-k_2)D(p_{s}-K)D(p_D-p_{s}+K)}. 
%\nonumber \\
\label{rt2}
\end{eqnarray}
First we integrate over $dK_+$ and $dk_{2+}$. To do this  we observe that only 
the poles in the  denominators  $D(p_s-k_2)$ and $D(p_s-K)$ give the finite 
contribution in the integral.
Other poles in $D(p_3+k_2)$ and $D(p_1+k_2-K)$  correspond to the  contribution 
at $|k_{2+}|\sim |p_{3+}|, |p_{1+}|\gg m$ and $K_+\sim p_{1+}\gg m$ and are strongly 
suppressed. Thus one  integrates by  $dK_+$ and $dk_{2+}$ by taking residues over 
the poles of the propagators $D(p_s-k_2)$ and $D(p_s-K)$ and obtains:
\begin{eqnarray}
A_{pd\rightarrow ppn} = {-1\over 4}\int {d^2K_\perp d^2k_{2\perp}\over (2\pi)^4}
 {dK_- dk_{2-}\over (2\pi)^2}  
{T_3(k_2)T_2(p_4-(p_D-p_{s}+K))T_1(K-k_2)\over 
D(p_3+k_2)D(p_1+k_2-K)}  \nonumber \\
{\Gamma_{DNN}\over (p_{s}-k_2)_{-} (p_{s}-K)_{-}D(p_D-p_{s}+K)}.
\label{rt3}
\end{eqnarray}
Next, to integrate over $dK_-$ and $dk_{2-}$ one observes that in high energy limit
when $p_{1+},p_{3+}\gg m$ and $p_{1-},p_{3-}\ll m$ the fast nucleon's propagator 
can be expressed as follows:
\begin{eqnarray}
-D(p_3+k_2) & = &  (p_3+k_2)_+(p_3+k_2)_- - (p_3+k_3)_\perp^2 - m^2 + i\epsilon \approx
p_{3+}(k_{2-} + i\epsilon) \nonumber \\ 
-D(p_1+k_2-K) & = & (p_1+k_2-K)_+(p_1+k_2-K)_- - (p_1+k_2-K)_\perp^2 - m^2 + i\epsilon 
\approx \nonumber \\
& & p_{1+}(k_{2-}-K_- + i\epsilon).
\label{prop}
\end{eqnarray}
Inserting these expressions into Eq.(\ref{rt3}) one obtains:
\begin{eqnarray}
A_{pd\rightarrow ppn} = {-1\over 4}\int {d^2K_\perp d^2k_{2\perp}\over (2\pi)^4}
 {dK_- dk_{2-}\over (2\pi)^2}  
{T_3(k_2)T_2(p_4-(p_D-p_{s}+K))T_1(K-k_2)\over 
p_{3+}(k_{2-}+i\epsilon)p_{1+}(k_{2-}-K_- +  i\epsilon)} \nonumber \\ 
{\Gamma_{DNN}\over (p_{s}-k_2)_{-} (p_{s}-K)_{-}D(p_D-p_{s}+K)}.
\label{rt3n}
\end{eqnarray}
Analyzing the range of the integration over  $dK_-$ and $dk_{2-}$ we first observe that 
the soft rescattering amplitudes depend predominantly on $q_\perp$, i.e.
$T_i(q)\sim T_i(q_\perp)$ and does not contain any 
singularities associated with $K_-$ and $k_{2-}$. Other apparent singularities may 
come from $(p_{s}-k_2)_{-}=0$ and $(p_{s}-K)_{-}=0$. However according to  the 
redefinitions of Eq.(\ref{momenta}) they correspond to $p_{s2-}=0$ and $p_{s1-}=0$. 
The latter conditions represent a bound nucleon with infinite virtuality 
$p_{s1+,2+}\sim {m^2\over p_{s1-,2-}}$ which is suppressed by wave function 
of bound nucleon. Therefore the structure of $dK_-$ and $dk_{2-}$ 
integrations will be defined only by two denominators of a fast propagating nucleon, 
i.e. by,
\begin{equation}
\int {d k_{2-} dK_-\over (k_{2-}+i\epsilon)(k_{2-}-K_- +  i\epsilon)} = 0.
\end{equation}
The above integral is zero since both poles over the $k_{2-}$ are on the same side 
of the complex $k_{2-}$ semi-plane and one can close the contour of integration 
on the side where there are no singularities exist. Thus 
this contribution results $A_{pd\rightarrow ppn} =0$.

This result allows us to reduce  potentially infinite number of rescattering diagrams 
to a finite set of diagrams. The only diagrams that survive are those in which a
propagating fast nucleon interacts first with one target nucleon then the next one 
and so on, making rescatterings strictly sequential. 

Another consequence of this theorem is that, if the virtuality of a bound nucleon, 
which is interacting  with the  propagating (energetic) nucleon, can be 
neglected, the sum of the interaction amplitudes with a given nucleon 
can be replaced by the invariant NN scattering amplitude, $F_{NN}$ as in Figure 11. 
The later can be replaced by the phenomenological NN scattering amplitude taken from 
the NN scattering data. 
\begin{figure}[h]  
\vspace{-0.4cm}  
\begin{center}  
\epsfig{angle=0,width=4.2in,file=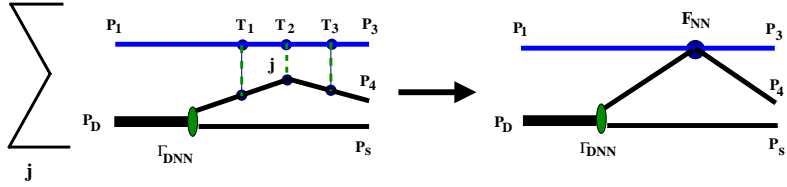}  
\end{center}  
\caption{The sum of the scattering vertices replaced by NN scattering amplitude.} 
\label{Fig.11}  
\end{figure}

Thus we will end up with the finite set of scattering diagrams for which 
Feynman diagram  rules can be identified.

The above result represents the realization of  eikonal approximation. However the 
major difference from the conventional semiclassical approximation is that the 
present approach does not require the spectator nucleons, to be a stationary 
scatterers\cite{Glauber}. Furthermore, we will refer 
the present approach as generalized eikonal approximation (GEA). 

\section{Feynman Diagram Rules for the Scattering Amplitude in GEA}
\label{FR}

In this section we  will define the effective Feynman diagram rules, within GEA, 
for the scattering amplitude of knocked-out  nucleon  to  undergo  $n$ rescatterings 
off the  nucleons of $(A-1)$ residual system. The case $n=0$ corresponds to the 
plane wave impulse approximation  in which the knocked out nucleon does not 
interact with residual nucleus. We systematically neglect the diffractive  
excitation of the nucleons in the intermediate states. In soft QCD  
processes this is a small correction for the knock-out nucleon (projectile) energies  
$\lesssim 10~GeV$. In the hard processes (that is when $Q^2$ - virtuality  
of the photon is sufficiently large ($\gtrsim 6-8~GeV^2$)) such an 
approximation can not be justified even  within this energy range, 
because of important role of quark-gluon degrees of freedom in  Color Transparency
phenomena, see for example discussion in  Ref.\cite{FMS94}. However our aim is  
to perform calculations in the kinematics where the color transparency phenomenon 
is still a small correction and intermediate hadronic states can be treated as a 
nucleon states.

\begin{figure}[h]  
\vspace{-0.4cm}  
\begin{center}  
\epsfig{angle=0,width=4.2in,file=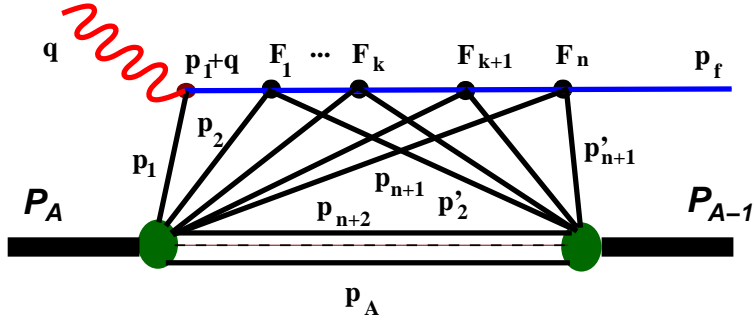}  
\end{center}  
\caption{ Diagram for n-fold rescattering.} 
\label{Fig.12}  
\end{figure}

According to the above discussion the n-fold rescattering amplitude 
will be represented through $n$ vertex amplitude in which the each vertex corresponds
to one NN scattering - Figure 12. We can formulate the following Feynman  rules of 
calculation of the diagram of Figure 12 (see also Ref.\cite{FSS97}).
\begin{itemize}
\item We assigns the vertex functions $\Gamma_{A}(p_1,...,p_A)$ and 
$\Gamma^\dag_{A-1}(p'_2,...,p_A)$ to describe the  transitions between 
$''nucleus~ A''$ to $''A~nucleons''$ with momenta $\{p_n\}$, 
$\{p'_n\}$ and $''(A-1)~nucleons''$ with momenta $\{p'_n\}$ 
to $``(A-1)~ nucleon~final ~state''$ respectively.

\item For $\gamma^* N$ interaction we assign vertex, $F^{em,\mu}_N$.
\item For each $NN$ interaction we assign the vertex function 
$F^{NN}_k(p_{k+1},p'_{k+1})$. This vertex function are related to 
the amplitude of $NN$ scattering as follows:
\begin{equation}
\bar u(p_3)\bar u(p_4)F^{NN}u(p_1)u(p_2) = 
\sqrt{s(s-4m^2)} f^{NN}(p_{3},p_{1})\delta_{\lambda,\lambda'} \approx 
sf^{NN}(p_{3},p_{1})\delta_{\lambda,\lambda'},
\label{NN}
\end{equation}
where $s$ is the total invariant energy of two interacting nucleons with momenta 
$p_1$ and $p_2$ and 
\begin{equation}
f^{NN} = \sigma_{tot}^{NN}(i+\alpha)e^{-{B\over 2}(p_3-p_1)_\perp^2},
\label{fnn}
\end{equation}
where $\sigma_{tot}^{NN}$, $\alpha$ and $B$ are known experimentally from $NN$ scattering 
data. The vertex functions are accompanied with $\delta$-function of energy-momentum 
conservation.

\item For each intermediate nucleon with four momentum $p$ we assign propagator 
$D(p)^{-1}=-(\hat p-m+i\epsilon)^{-1}$. Following to   
Ref.\cite{Gribov}  we choose the ``minus'' sign for the nucleon 
propagators to simplify the calculation of the overall sign of 
the scattering  amplitude.

\item The factor $n!(A-n-1)!$ accounts for  the combinatorics  
of $n$- rescatterings and $(A-n-1)$ spectator nucleons.

\item For each closed contour one gets 
the factor  ${1\over i(2\pi)^4}$   with no additional sign. 

\end{itemize}

Using above defined rules for the  scattering amplitude of Figure 12 
one obtains:
\begin{eqnarray}
& & F^{(n)}_{A,A-1}(q,p_f) = 
\sum\limits_{h}{1 \over n!(A-n-1)!
}\prod\limits_{i=1}^{A}
\prod\limits_{j=2}^{A} 
\int d^4p_i d^4p'_j 
{1\over \left[ i (2\pi)^4\right]^{A-2+n}}
\nonumber \\
& & 
\delta^4(\sum\limits_{i=1}^{A}p_i-{\cal P}_A) 
\delta^4(\sum\limits_{j=2}^{A}p'_j-{\cal P}_{A-1}) 
\prod\limits_{m=n+2}^{A}\delta^4(p_m-p'_m) \times 
\nonumber \\
& &  
{\Gamma_{A}(p_1,...,p_A)\over D(p_1)D(p_2)..D(p_{n+1})D(p_{n+2})..D(p_A)} 
{F^{em}_N(Q^2)\over  D(p_1+q)} {f^{NN}_1(p_2,p'_2)
  .. f^{NN}_n(p_{n+2},p'_{n+2}) 
\over D(l_1)..D(l_k).. D(l_{n-1})}
\times \nonumber \\       
& & {\Gamma_{A-1}(p'_2,.p'_{n+1},.p_{n+2}..,p_A)\over 
D(p'_2)..D(p'_{n+1})} 
\label{amp_n} 
\end{eqnarray}
where, for the sake of simplicity, we  neglect the spin dependent  
indices. Here  ${\cal P}_A$ and  ${\cal P}_{A-1}$ are the four  momenta 
of the target nucleus, and  final $(A-1)$ system, $p_j$ and $p'_j$  
are nucleon momenta in the nucleus $A$ and residual $(A-1)$ system 
respectively. $\sum\limits_{h}$ in Eq.(\ref{amp_n}) goes over virtual
photon  interactions  with different nucleons, where  $F^{em}_h(Q^2)$ 
are electromagnetic vertices. $-D(p_k)^{-1}$ 
is  the propagator  of a nucleon with momentum $p_k$ and $-D(l_k)^{-1}$ is 
the propagator of the struck nucleon in the intermediate  state, with momentum 
$l_k=q+p_1+\sum\limits_{i=2}^{k}(p_i-p'_i)$  between ${k-1}$-th and $k$-th  
rescatterings.  
The intermediate spectator states in the diagram of Figure 12
are expressed in terms of nucleons but not nuclear fragments
because the closure over  various nuclear excitations in the 
intermediate state is used. The possibility to use closure is related 
to the fact that the typical scale characteristic for high energy phenomena
is significantly larger then the energy scale of nuclear excitations
(for details see Appendix A).

After evaluation of the intermediate state nucleon propagators, the  
covariant amplitude  will be reduced to  a set of time ordered   
non covariant diagrams. This will help to  establish the correspondence
between the nuclear vertex functions and the nuclear wave functions. 
Particularly in the nonrelativistic limit the momentum space 
wave function is defined  through the vertex function as follows\cite{Gribov,Bert}:
\begin{equation}
\psi_{A}(p_1,p_2,...p_A)   =   {1\over (\sqrt{(2\pi)^3 2m})^{A-1}}
{\Gamma_A(p_1,p_2,...p_A)\over D(p_1)}, 
\label{wf_p} 
\end{equation}
where   wave functions are normalized as:   
$\int|\psi_A(p_1,p_2,...p_A)|^2 d^3p_1d^2p_2.. d^3p_A = 1$.

To demonstrate the application of the effective Feynman diagram rules we derive
formulae for the impulse approximation and first two rescattering   
terms (i.e. single and double rescatterings). To simplify  derivations 
we consider  $(e,e'N)$ reactions off the deuteron and $A=3$ target   
(see Figures 13,19 and 20)  and then  generalize the obtained results for 
the case of large $A$.

\section{Electro-disintegration of the Deuteron}

To demonstrate the application of the effective Feynman rules in the calculation of  
the nuclear scattering amplitude we first calculate the most simple case of 
electro-disintegration of the deuteron ($e+d\rightarrow e'+p+n$) in the kinematics 
of Eq.(\ref{kin}).  We will apply the following restrictions on the momenta involved 
in the reaction:
\begin{equation}
Q^2\ge 1~GeV^2, \ \ {\bf q}\approx {\bf p_f}\ge 1~GeV/c \  \mbox{and} \ 
|{\bf p_s}|\le 400 MeV/c
\label{kind}
\end{equation}
\begin{figure}[h]  
\vspace{-0.4cm}  
\begin{center}  
\epsfig{angle=0,width=4.2in,file=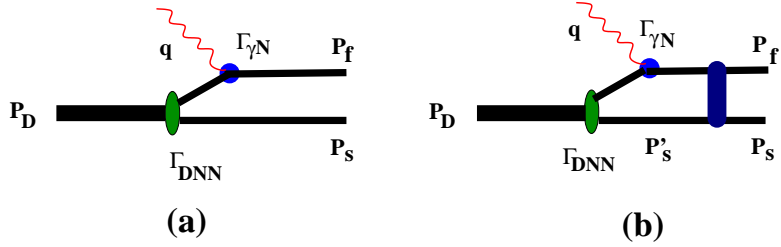}  
\end{center}  
\caption{Diagrams for $e+d\rightarrow e'+p+n$ reaction. (a) - PWIA contribution, 
        (b) - single rescattering contribution.} 
\label{Fig.13}  
\end{figure}
The last restriction allows us to neglect  by negative energy nucleon 
states (vacuum diagrams) and unambiguously identify the $D\rightarrow NN$ vertex 
multiplied by the propagator of virtual nucleon 
with the nonrelativistic wave function of the deuteron. Thus we will proceed with 
the calculation employing the approximation of ${\cal O}({p^2\over m^2})$
in calculation of virtuality of interacting nucleon and only nucleon 
degrees of freedom will be  taken into account. (Presence of negative energy states 
does not allow the above identification. The use of light-cone quantum mechanics 
helps to overcome this problem but as it was mentioned in Sec.IV the 
nuclear wave functions in this case are defined in the light-cone reference frame. 
However, in practice when  condition of (\ref{kind}) is satisfied all approaches 
should give rather close results (see e.g. \cite{FGMSS95}).)

\subsection{Plane Wave Impulse Approximation}

The amplitude corresponding to the impulse approximation diagram of Figure 13a can be 
written from Eq.(\ref{IA}):
\begin{equation}
A^\mu_0 =  -{\bar u(p_s)\bar u(p_f) \Gamma^{\mu}_{\gamma^*N}
\cdot [\hat{p_D}-\hat{p_s}+m]\cdot \Gamma_{DNN}\over (p_D-p_s)^2-m^2 + i\epsilon}.
\label{IAa1}
\end{equation}
The kinematic restriction on $p_s$ allows us to use approximation 
$\hat p_d-\hat p_s + m \approx \sum\limits_{\lambda}u_\lambda(p_d-p_s) 
\bar u_\lambda(p_D-p_s)$
and introduce electromagnetic current for bound nucleon 
\begin{equation}
j^\mu = \bar u(p_f)\Gamma^{\mu}_{\gamma^*N}u(p_D-p_s).
\label{elm}
\end{equation}
Note that this current still has an ambiguity related to the off-shellness of the bound 
nucleon however the effect is rather small for the kinematics of Eq.(\ref{kind}) 
(see e.g. \cite{FGMSS95}). Using the definition of $j^\mu$ from Eq.(\ref{elm}), 
for the scattering amplitude one obtains:
\begin{equation}
A^\mu_0 =  - {j^\mu \bar u(p_s)\bar u(p_D-p_s) \Gamma_{DNN}\over 
(p_D-p_s)^2-m^2 + i\epsilon}.
\label{IAa2}
\end{equation}
Next, because one neglects negative energy contributions we can identify the 
$D\rightarrow NN$ vertex with the nonrelativistic deuteron wave function as 
follows\cite{Gribov,BC,Bert}:
\begin{equation}
{\bar u(p_s)\bar u(p_D-p_s) \Gamma_{DNN}\over [m^2-(p_D-p_s)^2]
\sqrt{(2\pi)^32m}} = 
\psi_D(p_s)
\label{wfd}
\end{equation}
It can be proven that above defined wave function satisfies the nonrelativistic 
Schroedinger equation when kinematic condition given by Eq.(\ref{kind}) is valid.
Inserting Eq.(\ref{wfd}) into Eq.(\ref{IAa2}) for scattering amplitude within 
IA one obtains:
\begin{equation}
A^\mu_0 = \sqrt{(2\pi)^32E_s}\psi_D(p_s)j^\mu(p_s,q). 
\label{IAa3}
\end{equation}

\subsection{Single Rescattering Amplitude}

To calculate the amplitude corresponding to the single rescattering - Figure 13.b,  
we apply the Feynman rules of Sec.VII which results:
\begin{eqnarray}
A_{1}^\mu = -\int {d^4p'_s\over i (2\pi)^4}
{\bar u(p_f)\bar u(p_s)F_{NN}[\hat p'_s+m][\hat p_D-\hat p'_s + \hat q + m]\over
(p_D-p'_s + q)^2 - m^2 + i\epsilon}\nonumber \\
{\Gamma^{\mu}_{\gamma^*N}[\hat p_D-\hat{p'_s}+m]\Gamma_{DNN}\over 
((p_D-p'_s)^2-m^2 + i\epsilon)(p^{\prime 2}_s-m^2 + i\epsilon)}.
\label{F1_1}
\end{eqnarray}
One can integrate the above equation over $d^0p'_s$ observing that $F_{NN}$ does 
not have a singularity in $p'_s$. It reflects the fact that at high energies, the
total cross section of   $NN$ interaction  depends only weakly, 
on the collision energy (see e.g. \cite{Gribov,Bert}).

The kinematic restrictions of Eqs.(\ref{kin},\ref{kind}) (particularly 
$p_s\le 400 MeV/c$) allows us  to evaluate the loop integral in Eq.(\ref{F1_1}) 
by taking the residue over the spectator nucleon energy in  the intermediate state i.e. 
we can replace 
$[p'^2_s-m^2+i\epsilon]^{-1}d^0p'_s$ by $-i(2\pi)/2E'_s\approx -i(2\pi)/2m$. 
This is possible because in this case 
it is the only pole in the lower part of the $p'_{s0}$ complex plane. 
The calculation of the residue in  $p'_{s0}$  
fixes the time ordering from left to 
right in diagram Figure 13b.  Since this integration puts the spectator nucleon  in the 
intermediate state on its  mass-shell the $\Gamma_{DNN}$ vertex with the 
interacting (with photon) nucleon propagator will have  a similar construction as 
in the case of IA diagram. 
Thus using the relation of $\hat p_d-\hat p'_s + m 
\approx \sum\limits_{\lambda}u_\lambda(p_d-p'_s) \bar u_\lambda(p_d-p'_s)$ 
and $\hat p'_s + m  \approx \sum\limits_{\lambda}u_\lambda(p'_s) \bar u_\lambda(p'_s)$ 
one can  introduce the deuteron wave function according to Eq.(\ref{wfd}). 
Furthermore using Eqs.(\ref{NN},\ref{elm}) we obtain.
\begin{eqnarray}
& & A_{1}^\mu = \nonumber \\  
& & -{(2\pi)^{{3\over 2}}\sqrt{2E_s}\over 2m}\int {d^3p'_s\over  (2\pi)^3}
{sf_{pn}(p_{s\perp}-p'_{s\perp})\over (p_D-p'_s + q)^2 - m^2 + i\epsilon}
\cdot j^{\mu}_{\gamma^*N}(p_D-p'_s + q,p_D-p'_s)\cdot \psi_D(p'_s).
\label{F1_2}
\end{eqnarray}
Now we analyze the propagator of knocked-out nucleon:
\begin{equation}
(p_D-p'_s + q)^2 - m^2 + i\epsilon = m_D^2 - 2p_Dp'_s + p^{\prime 2}_s 
+ 2q(p_D-p'_s)-Q^2-m^2 + i\epsilon.
\label{knp1}
\end{equation}
We can simplify the Eq.(\ref{knp1}), using the relation of 
energy-momentum conservation:
\begin{equation}
(p_D - p_s+q)^2 = m^2 = m_D^2 - 2p_Dp_s + m^2+ 2q(p_D-p_s)-Q^2,
\label{rkn}
\end{equation}
which allows to represent Eq.(\ref{knp1}) as follows:
\begin{equation}
(p_D-p'_s + q)^2 - m^2 + i\epsilon = 
2|{\bf q}|\left[p'_{sz}-p_{sz} + {q_0\over |{\bf q}|}(E_s-m)
+ {m_D\over |{\bf q}|}(E_s-m) + {p^{\prime 2}_s-m^2\over 2|{\bf q}|}\right].
\label{knp2}
\end{equation}
Keeping only the terms which does not vanish with increase of $q$ (the 
first three terms in Eq.(\ref{knp2})) and observing that in the high energy 
limit (Eq.(\ref{small})) $2m|{\bf q}|\approx s$ for the scattering amplitude of 
Eq.(\ref{F1_2}) we obtain:
\begin{eqnarray}
A_{1}^\mu = 
 -{(2\pi)^{{3\over 2}}\sqrt{2E_s}\over 2}\int {d^3p'_s\over  (2\pi)^3}
{f_{pn}(p_{s\perp}-p'_{s\perp})\over 
p'_{sz}-p_{sz}+ \Delta + i\epsilon}
\cdot j^{\mu}_{\gamma^*N}(p_D-p'_s + q,p_D-p'_s)\cdot \psi_D(p'_s),
\label{F1_3}
\end{eqnarray}
where 
\begin{equation}
\Delta = {q_0\over |{\bf q}|}(E_s-m).
\label{Delta}
\end{equation}

The fact that the $f_{pn}$ depends only weakly  on 
the initial energy and is determined mainly by 
the transverse component of the  transferred momentum 
helps to carry out the integration over $p'_{sz}$ in Eq.(\ref{F1_3}).
For this one use the explicit form of the deuteron wave function 
as it is defined in the lab frame (see Eq.(\ref{eq.b13})). Inserting 
Eq.(\ref{eq.b13}) into Eq.(\ref{F1_3}) one can analytically integrate over
$p'_{sz}$ (see Appendix \ref{B}) arriving to the following expression 
for the single rescattering amplitude.
\begin{eqnarray}
A_{1}^\mu = 
 -{(2\pi)^{{3\over 2}}\sqrt{2E_s}\over 4i}\int {d^2k_t\over  (2\pi)^2}
f_{pn}(k_t)\cdot j^{\mu}_{\gamma^*N}(p_D-\tilde p_s + q,p_D-\tilde p_s)\cdot 
[\psi_D(\tilde p_s)-i\psi'_D(\tilde p_s)],
\label{F1_4}
\end{eqnarray}
where we defined the transverse component of the momentum transfered during $NN$ 
rescattering as $k_t = p'_{s\perp}-p_{s\perp}$. In Eq.(\ref{F1_4}) 
$\tilde p_s(\tilde p_{sz}, \tilde p_{s\perp}) \equiv 
{\bf \tilde p_s}(p_{sz}-\Delta, p_{s\perp}-k_\perp)$, 
$\psi_D$ is the deuteron wave function defined 
in Eq.(\ref{eq.b13}) and $\psi'_D$ is defined in
Eqs.(\ref{eq.b19}) and (\ref{eq.b20}).  Note that in general the $f_{pn}(k_t)$ which 
enters with $\psi_D$ and  $f_{pn}(k_t)$ that enters with $\psi'_D$ are different, 
since in the later case  $f_{pn}(k_t)$ corresponds to the off-shell 
amplitude\cite{FSP}. This off-shellness most likely will result to the reduction of 
the real part contribution.
However in many cases the overall contribution from $\psi'_D$ is small and off-shell 
effects in the  $f_{pn}(k_t)$ can be neglected.

\subsection{Relation of GEA to the Glauber Theory}

The next question we address is how the discussed above derivation 
of the rescattering amplitude is related to the Glauber theory.
To make a pointed comparison we first factorize the electromagnetic current 
in Eq.(\ref{F1_4}) from the integral. (The validity of such factorization will be 
discussed in the next Section.)  Furthermore  it is convenient to perform integration 
in Eq.(\ref{F1_3}) in coordinate space. Writing the Fourier transform of the deuteron 
wave function as:
\begin{equation}
\psi_D(p) = {1\over (2\pi)^{{3\over 2}}}\int d^3r \phi_D(r) e^{-ipr}, 
\label{eq.9}
\end{equation}
and using the coordinate space representation of the nucleon propagator:
\begin{equation}
{1\over [ p'_{sz}- p_{sz} + \Delta  + i\epsilon]}  = 
-i\int d z^0 \Theta(z^0) e^{ i (p'_{sz}- p_{sz} + \Delta )z^0},
\label{eq.10}
\end{equation}
we obtain  the formula for the rescattering amplitude:
 \begin{eqnarray}
A_1^\mu   &  = &  -j^\mu(p_s+q,p_s){\sqrt{2E_s}\over 2 i} 
\int{d^2 k_\perp\over (2\pi)^2}d^3r   
\phi(r)   f^{pn}(k_\perp)\theta(-z) e^{i(p_{sz}-\Delta)z} e^{i(p_{s\perp}-k_\perp)b} 
\nonumber \\ 
& = & -j^\mu(p_s+q,p_s){\sqrt{2E_s}\over 2 i} \int d^3r \phi(r) \theta(-z)\Gamma^{pn}
(\Delta,-z,-b) e^{ip_{s}r},
\label{eq.11}
\end{eqnarray}
where   $\vec r = \vec r_p - \vec r_n$ and 
we defined a generalized profile function $\Gamma$ as:
\begin{equation}
\Gamma^{pn}(\Delta,z,b) = {1\over 2 i}
e^{-i\Delta z}\int f^{pn}(k_\perp)e^{-ik_\perp b}  {d^2 k_\perp\over (2\pi)^2}.
\label{eq.12}
\end{equation}
One can see from Eq.(\ref{eq.12}) that Eq.(\ref{eq.11}) reduces to the Glauber
approximation in the limit of zero longitudinal momentum 
transfer $\Delta$. The dependence of the profile function 
on the longitudinal momentum transfer $\Delta$  originates from 
the nonzero  momentum of the recoiled nucleon, $p_s$. 
(Glauber theory is derived in the approximation of stationary nucleons, 
i.e. for zero momenta of spectator nucleons in the target.)
Even though, this new factor  in  
Eq.(\ref{Delta}), could be small it effects the cross 
section due to the steep momentum dependence 
of the deuteron wave function (see discussion in the next section).  
The same modified profile function, Eq.(\ref{eq.12}), 
is valid for the  single rescattering amplitudes of any nucleus 
$A$\cite{FSS97}. Thus one can conclude that in the limit of 
single rescattering the generalization of Glauber approximation to the GEA is 
rather simple, it requires the adding of phase factor - $\Delta$ in the Glauber 
profile function - Eq.(\ref{eq.12}). Note that this result is analogous to the 
account for the finite coherence length effects in the vector meson production 
from nuclei in eikonal approximation\cite{Yennie}.

\subsection{The Cross Section of Deuteron Electro-disintegration}

One can calculate now the cross section of the deuteron electro-disintegration
trough the electron and deuteron electromagnetic tensors as follows:
\begin{equation}
{d\sigma\over dE'_e d\Omega'_e d^3p_f/2E_f d^3p_s/2E_s} = 
{E'_e\over E_e} {\alpha^2\over q^4}
\eta_{\mu\nu} T^{\mu\nu}_D \delta^4(p_D+q-p_f-p_s),
\label{sigma}
\end{equation}
where $\eta_{\mu\nu} = {1\over 2}Tr(\hat k_2\gamma_\mu \hat k_1\gamma_\nu)$. 
$k_1\equiv (E_e, \vec k_1)$ and $k_2\equiv (E'_e, \vec k_2)$ are 
the four-momenta of incident and scattered electrons respectively. 
The electromagnetic tensor  $T_D^{\mu\nu}$ of the  deuteron is: 
\begin{equation}
T_D^{\mu\nu} =\sum\limits_{spin} (A_0+A_1)^\mu(A_0+A_1)^\nu,
\label{Ttens}
\end{equation}
where $A_0$ and $A_1$ correspond to impulse approximation and single rescattering 
amplitudes calculated in Sec.VIII~(A) and (B). 
For the  numerical calculation we apply the factorization approximation 
in calculating the rescattering amplitude of Eq.(\ref{F1_4}). Namely 
using the fact that in soft $NN$ rescattering 
$<k^2_{\perp}>_{rms}\sim 250 MeV^2/c^2 \ll |{\bf q}|^2$ the bound nucleon 
electromagnetic current can be factorized out of the integral in 
Eq.(\ref{F1_4}). In this case one arrives to the distorted wave impulse 
approximation (DWIA), in which the scattering cross section could be 
represented as a product of the off-shell $eN$ scattering cross section-$\sigma_{eN}$ 
and the distorted spectral function-$S_D(p_f,p_s)$.
\begin{equation}
{d\sigma\over dE'_e d\Omega'_e dp_f} = p_f^2\sigma_{eN}S_D(p_f,p_s).
\label{dwia}
\end{equation}
The off-shell cross section $\sigma_{eN}$ contains ambiguity in the spinor part
related to the fact that knocked-out nucleon is bound (see e.g. \cite{deFor}). 
However the restrictions of Eq.(\ref{kind}) keep these ambiguities 
for $q\sim few~GeV/c$ region reasonably small (see Ref.\cite{FGMSS95}).

The distorted spectral function can be represented as follows
\begin{equation}
S(p_f,p_s) = \left|\psi_D(p_s) -{1\over 4i}\int {d^2k_t\over  (2\pi)^2}
f_{pn}(k_t)\cdot [\psi_D(\tilde p_s)-i\psi'_D(\tilde p_s)]\right|^2.
\label{sf}
\end{equation}

To analyze the effects of rescattering in the cross section we 
calculate the ratio of the cross section of Eq.(\ref{dwia}) to the 
cross section calculated within plane wave impulse approximation, in 
which only IA amplitude~-~$A_0$ is included:
\begin{equation}
T = {\sigma^{IA+FSI}\over \sigma^{IA}} = {S(p_f,p_s)\over |\psi_D(p_s)|^2}.
\label{T}
\end{equation}

Figure 14 demonstrates the calculation of $T$ as a function of the 
recoil nucleon angle $\theta_{sq}$ with respect to the ${\bf q}$ for the different 
values of recoil nucleon momentum.
The Figure 14 demonstrates the distinctive angular dependence of ratio T.
At recoil nucleon momenta $p_s\le 300 MeV/c$,  $T$ has a minimum and generally 
$T<1$ while  at $p_s >300 MeV/c$, $T >1$ and  has a distinctive maximum. 

One can easily understand the structure of $T$ if  to recall that 
the soft rescattering amplitude is mainly imaginary 
$f_{pn}= \sigma_{tot}(i+\alpha)e^{-{B\over 2}k^2_\perp}$ with $\alpha\ll 1$. 
In this case inserting Eq.(\ref{sf}) into Eq.(\ref{T}) one obtains 
for $T$:
\begin{eqnarray}
T \approx 1 - {1\over 2}\left|{\psi_D(p_s)\cdot \int{d^2k_\perp\over 
(2\pi)^2}f_{pn}(k_\perp)\cdot 
[\psi_D(\tilde p_s)-i\psi'_D(\tilde p_s)]\over \psi^2_D(p_s)}\right| + \nonumber \\
{1\over 4}{\left| 
\int{d^2k_\perp\over (2\pi)^2}f_{pn}(k_\perp)
\cdot[\psi_D(\tilde p_s)-i\psi'_D(\tilde p_s)]\right|^2
\over  \psi^2_D(p_s)}.
\label{T_2}
\end{eqnarray}

\begin{figure}[th]  
\vspace{-0.6cm}  
\begin{center}  
\epsfig{angle=0,width=4.2in,file=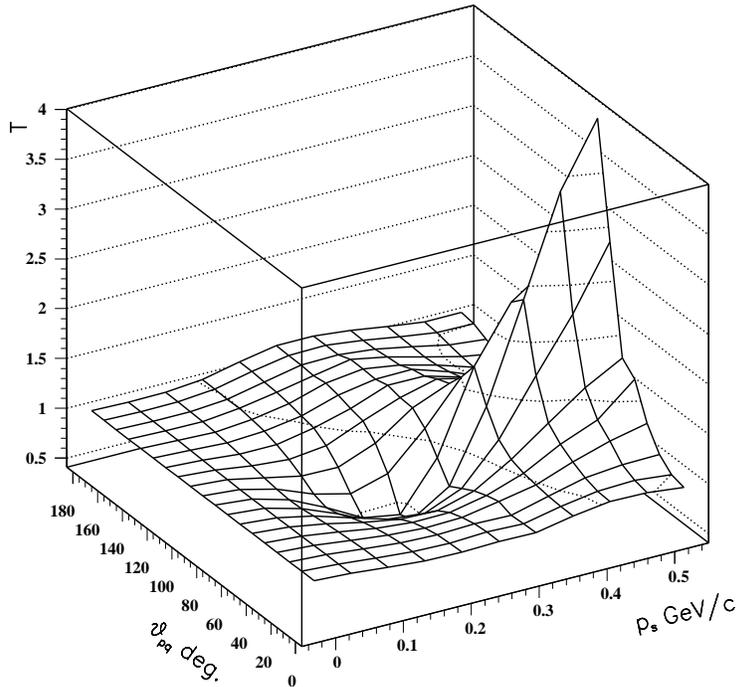}  
\end{center}  
\caption{The dependence of the transparency T on the angle, $\theta_{sq}$  
and the  momentum, $p_s$ of the recoil nucleon. The angle is defined with 
respect to the ${\bf q}$.} 
\label{Fig.14}  
\end{figure}

From Eq.(\ref{T_2}) one observes that the interference term has a negative sign and 
is proportional to ${\psi_d(p_{sz}, p_{s\perp})\psi_d(p_{sz}-\Delta, p_{s\perp}-k_\perp)
\over\psi_d^2(p_{sz},p_{s\perp})}$. 
Thus in the kinematics when the interference term is dominant the $pn$ rescattering 
results in the screening of the overall cross section, thus $T<1$. The maximal 
screening is found at $p_{st}\approx 200 MeV/c$ at which the square of rescattering term
(third term in Eq.(\ref{T_2})) is small and $T\le 1$. Further increase of $p_s$ 
suppresses the relative contribution of interference term as compared to the square of 
rescattering term  which results to $T>1$. The dominance of 
the rescattering term as compared to the interference term, with increase of $p_s$, 
can be understood from the fact that the  interference term grows as 
$\sim 1/|\psi_D(p_s)|$ while rescattering term $\sim 1/|\psi_D(p_s)|^2$.

\subsection{Once More about the Relation of GEA to the Glauber Theory}

It is very illustrative to compare the predictions for $T$ calculated within GEA and 
Glauber approximation.

As figure 15 demonstrates the predictions based on GEA are rather close to that for 
the Glauber approximation when recoil nucleon momentum is small. 
(Recall that Glauber theory is derived for the 
cases  when target nucleons are considered as a stationary scatterer and therefore 
their Fermi momenta have been neglected.) However at larger Fermi momenta 
predictions of both approaches differ considerably. For example for 
$p_s=400~MeV/c$ the GEA and Glauber 
approximation predictions for  angular dependence of the maximal contribution of  
the rescattering amplitude (i.e. the position of the maximum of $T$ in Figure 15) 
differ by as much as  $30^0$. Such a difference is quite dramatic and can be checked 
in the forthcoming experiments at Jefferson Laboratory\cite{KG,EGM,WKV,UJ}.

\subsection{The $Q^2$ Dependence of $T$ and Phenomenon of Color Coherence}

So far we discussed only the $p_s$ and $\theta_{p_sq}$ dependence of $T$. 
However another interesting feature of $T$ is its $Q^2$ dependence. 
As it was discussed previously one of the important features of high energy 
scattering is the energy independence of $NN$ soft scattering cross section 
(e.g. Figure 8), which enters into the rescattering amplitude 
(Figure 13b) of the $d(e,e'N)N$  reaction. According to Eqs.(\ref{kind}) and 
(\ref{fnn}) such  energy independence should be reflected in the $Q^2$ independence 
of $T$ (Eq.\ref{T_2}) at fixed values of recoil nucleon momenta. 
As Figure 16 shows indeed with increase of $Q^2$, $T$ becomes practically $Q^2$ 
independent at given value of $p_s$. 

\begin{figure}[h]  
\vspace{-0.6cm}  
\begin{center}  
\epsfig{angle=0,width=4.2in,file=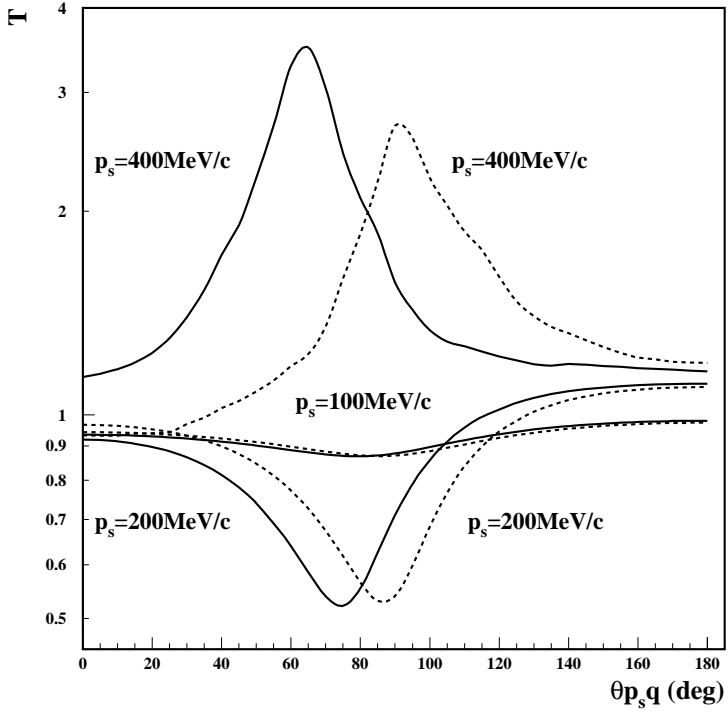}  
\end{center}  
\caption{The $\theta_{p_sq}$ dependence of T at different values of $p_s$. 
Solid lines- GEA, dashed lines - Glauber approximation.} 
\label{Fig.15}  
\end{figure}

The $Q^2$ independence of $T$  within GEA can be used as a baseline reference point
to study the onset of the color coherence regime in high $Q^2$ exclusive 
reactions off nuclei. The idea of this phenomenon is based on the observation that at 
high $Q^2\ge Q^2_0$ ($Q^2_0\approx 6-8 GeV^2$) the elastic 
$\gamma^*-N$ interaction is  dominated by the contribution of minimal Fock component of 
quark-gluon wave function of the nucleon.  The minimal Fock component of the hadronic 
wave function corresponds to the small sized - point like configurations (PLC) in the 
hadrons. Thus at  $Q^2\ge Q^2_0$ in the QCD picture one expects that hard elastic 
scattering will select the PLC from the wave function of the nucleon. Note that this 
feature is characteristic of both the  perturbative\cite{BF,Mueller} and 
nonperturbative\cite{FMS93} regime of QCD. 
If PLC are indeed produced in   high $Q^2$ exclusive reactions then because of 
color screening (PLC are color singlet objects) one  expects that they should propagate 
through the nuclear medium without final state interactions\cite{B82,M82}- a 
phenomenon called Color Transparency (CT) (for review see Refs.\cite{FMS94,JPR}). 
This phenomenon is observed at FNAL in high energy regime when perturbative
QCD is valid (see Refs.\cite{BFS,FMS93b,Ashery} and reference therein). 

However in the nonperturbative QCD domain, observation of color coherence 
phenomena is complicated by the fact that at finite energies PLC will evolve to 
the normal hadronic state during its propagation through the nucleus.
\begin{figure}[h]  
\vspace{-0.4cm}  
\begin{center}  
\epsfig{angle=0,width=4.2in,file=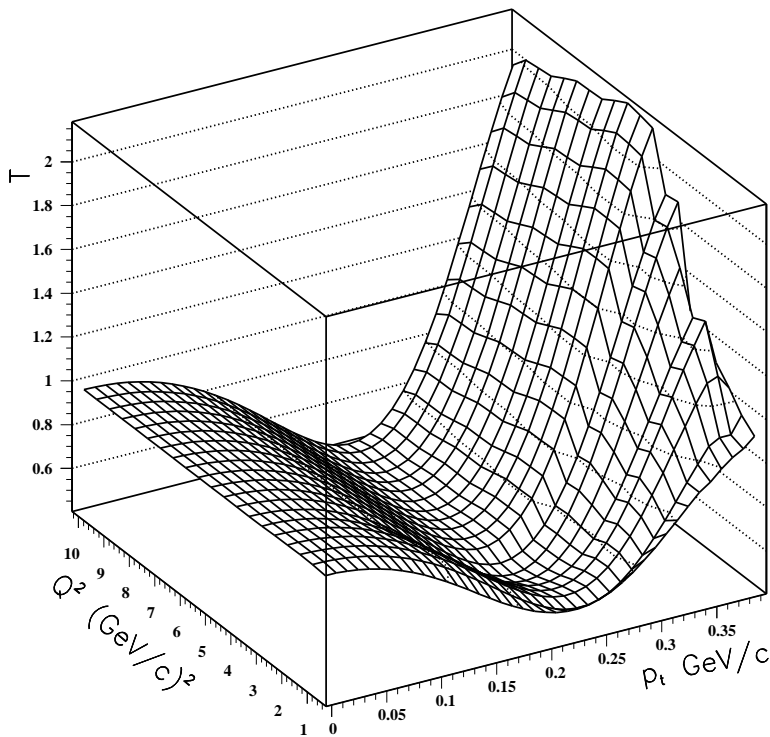}  
\end{center}  
\caption{$T$ as a function of $Q^2$ and $p_{st}$ at $\alpha_s = {E_s-p_{sz}\over m}=1$.} 
\label{Fig.16}  
\end{figure}
As a result  
FSI will not be negligible at the later stages of the interaction.  It is believed 
that the expansion of PLC was the major reason that prevented  the observation of 
the onset of color coherence in $A(e,e'p)X$ reactions at 
the range of $Q^2 \sim 8 GeV^2$\cite{NE18,Ent}. Thus the major issue in the studies of 
color coherence phenomena in the nonperturbative regime is the suppression of the 
expansion of PLC.  As was shown in Ref.\cite{FGMSS95,FGMSS96} such a suppression 
can be achieved in exclusive $d(e,e'N)N$ reactions in which the selection of larger 
values of $p_{s\perp}$ allows the suppression of the distance between the space-time 
point of $\gamma^*N$ interaction  and $PLC-N$ reinteraction. As a result one can 
considerably suppress the expected effects of a PLC expansion. 

There are several nonperturbative models which allow an estimation of the expected 
effects of color coherence incorporating a PLC expansion 
(see e.g. \cite{FFLS,MJ,FGMS,Nikol,Kopel,BCT}).  To illustrate the expected magnitude 
of the phenomenon we use the Quantum Diffusion Model~(QDM) of Ref.\cite{FFLS}.  
In QDM  the reduced interaction  between the PLC and the spectator neutron 
(to be specific, we use a neutron as a spectator)  can be 
accounted for by introducing the dependence of the  scattering amplitude on 
the transverse size of the PLC. However since we  consider  energies that are far from 
asymptotic, the expansion of PLC  should be important. This feature is 
included by allowing the rescattering amplitude to depend on the distance from 
the photon absorption point. Within QDM for PLC-N scattering amplitude one obtains 
\cite{FFLS,FGMSS95}:
\begin{equation}
f^{PLC,N}(z,k_t,Q^2) = i\sigma_{tot}(z,Q^{2}) \cdot 
e^{{b\over 2 }t}\cdot {G_{N}(t\cdot\sigma_{tot}(z,Q^{2})/\sigma_{tot})
\over G_{N}(t)},  
\label{F_NNCT}  
\end{equation}  
where $b/2$ is the slope of elastic $NN$ amplitude,    $G_{N}(t)$ 
($\approx  (1-t/0.71)^{2}$) is the Sachs form factor and 
$t= -k_t^2 $. The last factor in Eq.(\ref{F_NNCT}) accounts for
the difference  between elastic  scattering of  PLC and average
configurations, using  the observation that the $t$ dependence
of $d\sigma^{h+N\rightarrow h+N}/dt $ is roughly that of 
$\sim~G_{h}^{2}(t)\cdot G_{N}^{2}(t)$ for not very large values of t 
and that $G_{h}^{2}(t)\approx exp(R_h^2t/3)$.

In Eq.~(\ref{F_NNCT}) $\sigma_{tot}(l,Q^{2})$  is the  effective total 
cross section of the  interaction  of the PLC at the distance $l$ from 
the interaction point. The quantum diffusion   model~\cite{FFLS}
corresponds to:
\begin{equation}
\sigma _{tot}(l,Q^{2}) = \sigma_{tot} \left \{ \left ({l \over l_{h}} + 
{\langle r_{t}(Q^2)^{2} \rangle \over \langle r_t^{2}  \rangle } 
(1-{l \over l_{h}}) \right )\Theta (l_{h}-l) + \Theta (l-l_{h})\right\}, 
\label{SIGMA_CT}  
\end{equation} 
where ${l_h = 2p_{f}/\Delta~M^{2}}$, with ${\Delta~M^{2}=0.7-1.1~GeV^{2}}$. 
Here ${\langle r_{t}(Q^2)^{2} \rangle}$  is the average transverse size 
squared of the  configuration  produced at the interaction  point.  
In several realistic models considered  in Ref.\cite{FMS92} it can be 
approximated as ${ {\langle r_{t}(Q^2)^2\rangle\over\langle r_t^2\rangle} 
\sim{1\,GeV^2\over Q^2}}$ for  $Q^2~\geq~1.5~GeV^2$.  Note that due to 
effects of expansion the results of calculations are rather insensitive 
to the value of this ratio whenever it is much less than unity. The calculation 
of the deuteron electrodisintegration cross section  within the QDM is performed 
by rewriting the amplitude of Eq.(\ref{F1_4}) in coordinate representation 
and using $f^{PLC,N}$ from Eq.(\ref{F_NNCT}) to replace the amplitude of 
$NN$ scattering, $f_{pn}$. 

Other nonperturbative models which incorporate both the production of PLC and its 
expansion during the propagation based on the observation that with increase of 
energies the contribution from the inelastic transitions with intermediate 
baryonic resonances of the same spin as the nucleon (as $N^*$ and $N^{**}$) are 
not further suppressed. As a result one expect that the intermediate hadronic 
state (after the $\gamma^*N$ vertex in the diagram of Figure 13b) to represent 
the superposition of the  nucleonic, baryonic excitation and continuum states.
The requirement that  at high $Q^2$ $\gamma^*N$ produces a PLC imposes 
an additional constraints on the structure of this intermediate state.

In practical calculations the intermediate state is modeled 
through the two or  three resonance states\cite{MJ,FGMS,Kopel,BCT}. 
The three-state model is based on the assumption that the hard  $\gamma^*N$ 
scattering operator produces a non-interacting PLC which is a superposition of three 
baryonic states\cite{FGMS}:
\begin{equation}
|PLC> = \sum\limits_{m=N,N^{*},N^{**}} F_{m,N}(Q^2)|m>,
\label{3s}
\end{equation}
where $F_{m,N}(Q^2)$ are elastic ($m=N$) and inelastic transition form factors. 
The non-interaction of the initially produced PLC is provided by the condition:
$T_S|PLC>=0$, where $T_S$ is the  $3\times 3$ Hermitian matrix representing 
the small angle final state interactions. The relevant cross section is 
obtained from Eq.(\ref{F_NNCT}) replacing $f_{pn}$ by $T_S$.
The detailed numerical calculations done in Ref.\cite{FGMSS95} demonstrated 
the both QDM and resonance models predict the similar magnitudes for the 
effect of the suppression of final state rescatterings. Thus in further 
estimations we will be restricted by QDM calculations only.

To be able to register the signature of color coherence effect it is important to 
identify the quantity which is most sensitive to the suppression of FSI with increase of 
$Q^2$. As can be seen from Figure 14  within GEA the FSI is dominant 
in the transverse kinematics $\theta_{p_sq}\approx 90^0$ (more precisely when 
$\alpha_s = {E_s - p_{sz}\over m} = 1$). Besides as follows from Figures 14 and 15 at 
$p_{s}\lesssim 300 MeV/c$  FSI dominates in the interference term of Eq.(\ref{T_2}) thus 
the screening term is dominant and  $T<1$. 
At $p_{s}\gtrsim 350 MeV/c$ the dominant contributions arises from the double 
scattering term in Eq.(\ref{T_2}) and as a result $T>1$. This observation allows to 
define  the ratio of the cross section, measured at  kinematics where  
double scattering is dominant, to the cross section measured at 
kinematics where the effect of the  screening is more
important. 

\begin{figure}[h]  
\vspace{-0.4cm}  
\begin{center}  
\epsfig{angle=0,width=4.2in,file=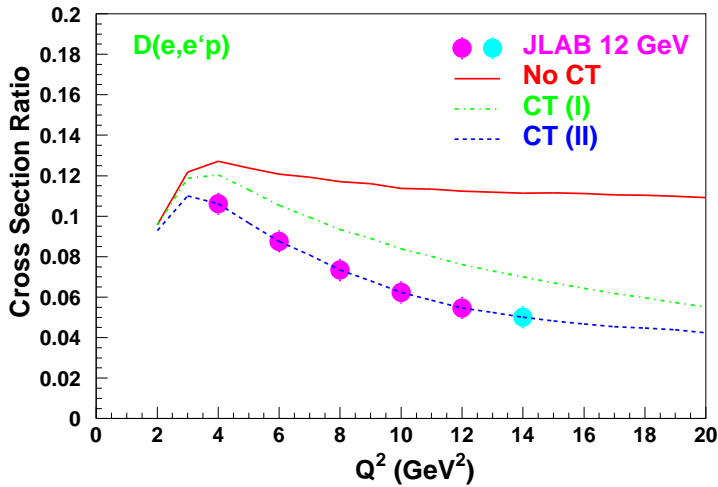}  
\end{center}  
\caption{The $Q^2$ dependence of the ratio $R$ as defined in Eq.(\ref{rdouble}) for 
the reaction $d(e,e'p)n$ reaction. The solid curve is generalized eikonal approximation.
Dashed and dash-doted curves correspond to the prediction of quantum diffusion model 
with $\Delta M=0.7 GeV^2$ (maximal CT) and  $\Delta M=1.1 GeV^2$ (minimal CT) 
respectively. The full circles are the projected data for 12 GeV upgrade of 
JLAB~[14].}
%\cite{wp} }
\label{Fig.17}  
\end{figure}
  
The Figures 14  and 15 shows that it is possible  to
separate these two kinematic regions  by choosing two momentum intervals for    the 
recoil nucleon,   (350--500~MeV/c) for double scattering and    (0--250 MeV/c) for the 
screening.   Thus the suggested experiment will measure  the
$Q^2$ dependence of    the following typical ratio:
  \begin{equation}  
R = {\sigma(p =400MeV/c)\over \sigma(p=200~MeVc)}.
\label{rdouble}  
\end{equation}  
As figure 17 demonstrates the onset of color transparency will result the decrease of the 
ratio $R$ with increase of $Q^2$, while GEA predicts practically no $Q^2$ dependence 
starting at $Q^2\ge 4GeV^2$. The circles in the Figure 17 correspond to the 
experimental data projected for the 12 GeV upgrade of CEBAF at Jefferson Lab\cite{wp}.
Note that, in addition to the d(e,e'pn) process, one can consider excitation of baryon
resonances  with spectator kinematics, like $D(e,e'pN^*)$ and  
$D(e,e'N\Delta)$.   The later process is of special  interest for looking
for  the so called chiral    transparency, disappearance of the pion field of the 
ejectile~\cite{chiral,FGMSS96}.  
   
\subsection{Relation to the Calculations Based on Intermediate Energy Approaches}

For completeness of the discussion on electro-disintegration of the deuteron we  
address the question about overlap and continuity between the predictions of 
approaches of intermediate and high 
energy physics. 
\begin{figure}[h]  
\vspace{-0.7cm}  
\begin{center}  
\epsfig{angle=0,width=4.0in,file=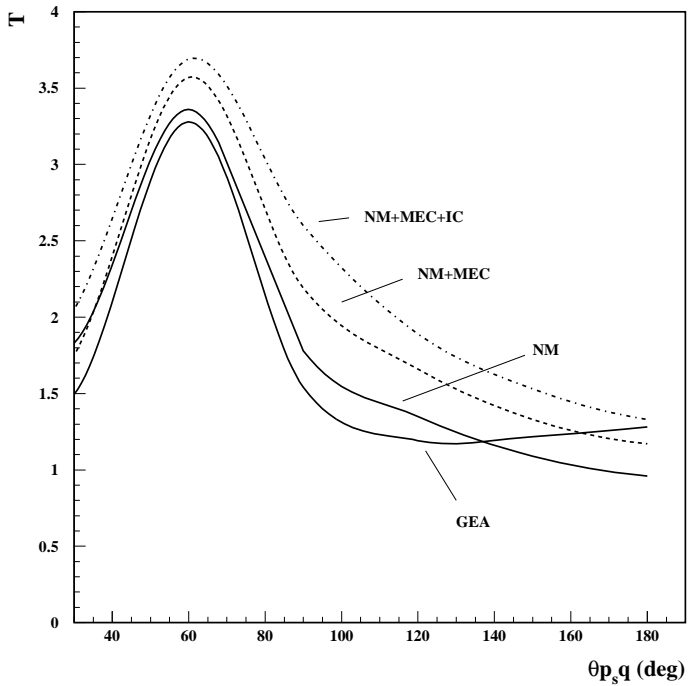}  
\end{center}  
\caption{The $\theta_n$ dependence of ratio T at $Q^2=1~GeV^2$. 
The curve marked by ``NM'' - corresponds to the cross section including  
final state interactions calculated  in Ref.~[44].
%\cite{Arenhovel}. 
The curves with ``NM+MEC'' and ``NM+MEC+IC'' 
correspond to the calculations of Ref.~[44]
%\cite{Arenhovel}  
including meson exchange currents  and Isobar contributions  respectively.} 
\label{Fig.18}  
\end{figure}

Practically important question is  whether the predictions of  intermediate 
and high energy theories match at borderline kinematics. 
As an borderline kinematics we consider the electro-disintegration of the 
deuteron at $Q^2=1~GeV^2$ with all other kinematical 
constraints defined in Eq.(\ref{kind}). In figure 18 we compare the prediction GEA with 
the results of calculations of the model of Ref.\cite{Arenhovel}. Here $Q^2=1~GeV^2$ 
can be considered as an upper limit for the model of Ref.\cite{Arenhovel}, in which up 
to seven orbital moments are included to account for the final state reinteraction. 
On the other hand $Q^2=1~GeV^2$ corresponds to  the lower limit of  validity of GEA 
since the  kinematical constrains of Eq.(\ref{small}) is barely satisfied.
As figure 18 shows one has surprisingly good  agreement between GEA and model of 
Ref.\cite{Arenhovel} when no  MEC and IC contributions are included. 
As we discussed above with increase of $Q^2$ relative contribution of MEC and IC should 
decrease as compared with IA and FSI. This demonstrates 
the existence of continuity between both methods of calculations. 

\section{Electro-disintegration of Nuclei with $A\ge 3$}

We consider now the electro-disintegration of nuclei with $A\ge 3$ (Figure \ref{Fig.19}). 
The impulse approximation and single rescattering contribution can be generalized from 
the Eq.(\ref{IAa3}) and (\ref{F1_3}) respectively.

\begin{figure}[h]  
\vspace{-0.4cm}  
\begin{center}  
\epsfig{angle=0,width=4.2in,file=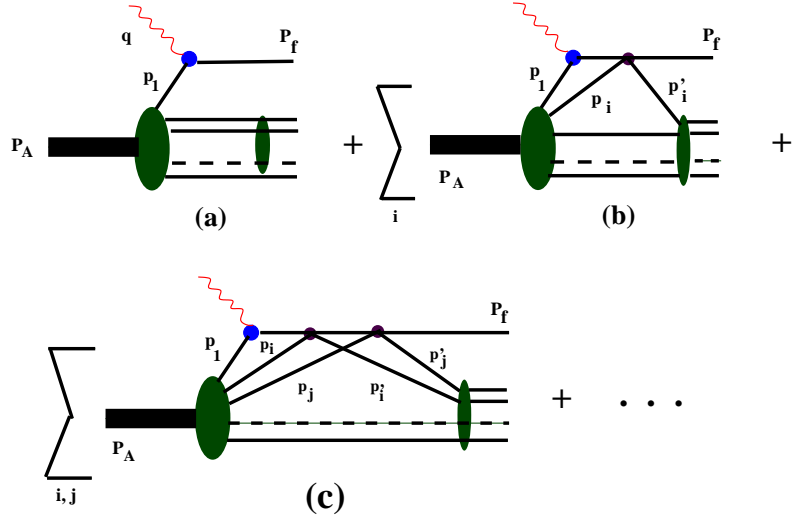}  
\end{center}  
\caption{Diagrams of A(e,e'N)X reactions. (a) - PWIA term, (b) - single rescattering 
term, (c) - double rescattering term, etc.} 
\label{Fig.19}  
\end{figure}

\subsection{Impulse Approximation} 
For IA term one obtains:
\begin{equation}
A_0^\mu = F\int \Psi^+_{A-1}(p_2,p_3,...,p_A)j^\mu(p_m,q)
\Psi_A(p_m,p_2,p_3,...,p_A)d^3p_2 d^3p_3 ... d^3p_A, 
\label{IAm}
\end{equation}
where ${\bf p_m} = {\bf p_f} - {\bf q}$. 
$F$ - is the phase space factor for the residual $A-1$ nuclear system. For example, 
for two-body break-up reactions $F=\sqrt{(2\pi)^3 2E_{A-1}}$ or for tree-body break-up 
reaction on $A=3$ target $F=\sqrt{(2\pi)^32E_{s1}(2\pi)^32E_{s2}}$ in which $E_{s1}$ and 
$E_{s2}$ are the energies of two recoil nucleons. 

\subsection{Single Rescattering Contribution}

For the single rescattering term (Figure \ref{Fig.19}(b)) we generalize Eq.(\ref{F1_3}) 
to obtain:
\begin{eqnarray}
A_1^\mu  & = &   -{F\over 2}\sum\limits_{i=2,A}\int \Psi^+_{A-1}(p_2,p'_i,...,p_A)
{f_{NN}((p'_i-p_i)_\perp)\over p_{mz}-p_{1z}+\Delta + i\epsilon}
j^\mu(p_1,q) \Psi_A(p_1,p_2,p_i,...,p_A)\nonumber \\
& &  \times {d^3p_1\over (2\pi)^3} d^3p_2, ...'...d^3p_A,
\label{F1m}
\end{eqnarray}
where ``$^\prime$'' in the integration means it does not contain $p_i$. From 
energy-momentum conservation one has $p_1 = {\cal P}_A - p_i - p_2 - p_A$ and 
$p'_i = {\cal P}_{A-1}-p_2-...'...p_A$, 
where  ``$^\prime$'' means the exclusion of $i$-th nucleon's momentum.

If we transform the integrations into coordinate space one can see easily that the only 
difference from the Glauber theory is the modification of the profile function according 
to Eq.(\ref{eq.12}). Therefore the relation between GEA and Glauber theory in the limit 
of single rescatterings is rather straightforward and the only modification required is 
the phase factor in Eq.(\ref{eq.12}). Note that the $\Delta$ factor depends on the 
particular reactions. 
For example in the case of a two-body break-up 
\begin{equation}
\Delta={q_0\over |{\bf q}|}{p_m^2\over 2M_{A-1}} + |\epsilon_b|,
\label{Delta2}
\end{equation}
were $\epsilon_b$ is the nuclear binding energy. For this relation one can see that 
for the case of large $A$, the  $\Delta\sim 0$ for large values of $p_m$. Therefore one can 
conclude that the Glauber theory can  be extended to the region of large values of 
missing-momenta $p_m$ for two-body break-up reactions with large $A$ nuclei\cite{FMSS95}. 
For the case of three-body break-up of a $A=3$ target:
\begin{equation}
\Delta = {q_0\over |{\bf q}|}(T_{s1} + T_{s2} + |\epsilon|),
\label{Delta3}
\end{equation}
were $T_{s1}$ and $T_{s2}$ is the kinetic energy of the two recoil nucleons in the 
reaction and it can not be neglected in the processes with large values of missing 
momentum $p_m$.

\subsection{Double Rescattering Contribution}

For double rescattering contribution we have to calculate the two diagrams of
Figure 20. Using the Feynman diagram rules from Section VII for 
amplitude of Figure 20a one obtains:
\begin{figure}[h]  
\vspace{-0.4cm}  
\begin{center}  
\epsfig{angle=0,width=6.4in,file=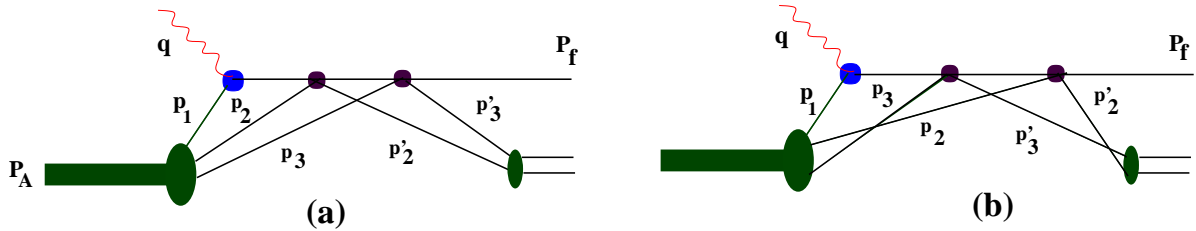}  
\end{center}  
\caption{Double rescattering diagram for $^3He(e,e'N)X$ reaction.}  
\label{Fig.20}  
\end{figure}

\begin{eqnarray}  
& & A_{2a}^\mu   =  \nonumber  \\
& & \int  {\Gamma^+(p'_2,p'_3)\over  D(p'_2) D(p'_3)}
{F^{NN}(p'_3-p_3)\over D(p_1+q+p_2-p'_2)}
{F^{NN}(p'_2-p_2)\over D(p_1+q)}  j^\mu(p_1,q)
{\Gamma(p_1,p_2,p_3)\over D(p_1) D(p_2) D(p_3)} 
\nonumber \\
& & \delta^4(p_A- p_2 - p_3 - p_1) 
 \delta^4(p_{A-1}- p'_2 - p'_3) d^4p_1 d^4p_2 d^4p_3  d^4p'_2 d^4p'_3 
\left[{1\over i (2\pi)^4}\right]^3
=  \nonumber \\ 
& & \int 
{\Gamma^+(p'_2,p'_3)\over  D(p'_2)D(p'_3)} 
{f^{NN}(p'_3-p_3)\over D(p_1+q+p_2-p'_2)} {f^{NN}(p'_2-p_2)\over D(p_1+q)}
\times \nonumber \\
& & \ \ \ \ \ \ \ \ j^\mu(p_1,q) {\Gamma(p_1,p_2,p_3)\over D(p_1) D(p_2) D(p_3)}
{d^4p_2\over i(2\pi)^4} {d^4p_3\over i(2\pi)^4}{d^4p'_3 \over i(2\pi)^4},
 \label{eq.25}
\end{eqnarray}
where
\begin{equation}
p_1 = {\cal P}_{A} - p_3 - p_2; \ \ \ \ \ \ \ 
p'_2 = {\cal P}_{A-1} - p'_3.
\label{eq.26}
\end{equation}
Here again for simplicity we are neglected the spins.
Then, using the same approximations as for the cases of IA and single 
rescattering amplitudes we can perform integration over $d^0p_2$, $d^0p_3$, 
$d^0p'_3$, which effectively results in the replacement 
$\int{d^0p_{j}\over 2\pi i D(p_{j})} 
\rightarrow  {1\over 2 E_{j}} \approx {1\over 2m}$, ($j=2,3,3'$).

Using  Eq.(\ref{eq.26}) and the definition of the initial and final state 
wave functions from the relations analogous to the Eq.(\ref{wfd}) we obtain:
\begin{eqnarray}
A^\mu_{2a} & = &  {F \over 4m^2} 
\int \psi^+_{A-1}(p'_2,p'_3)
{F^{NN}(p'_3-p_3)\over D(p_1+q+p_2-p'_2)}
{F^{NN}(p'_2-p_2)\over D(p_1+q)} \times \nonumber \\ 
&  &  j^\mu(p_1,q) \psi_A(p_1,p_2,p_3)
{d^3p_1\over (2\pi)^3}{d^3p_3\over (2\pi)^3}{d^3p'_3\over (2\pi)^3},
\label{eq.27}
\end{eqnarray}
where $D(p_1+q)$ is the same as for the 
case of single rescattering. In high energy limit:
\begin{equation}
- D(p_1+q)\approx  2|{\bf q}|(p_{zm}-p_{1z} + \Delta + i\epsilon),
\end{equation}
were $\Delta$ is defined by Eq.(\ref{Delta2}) or (\ref{Delta3}).
For $D(p_1+q+p_2-p'_2)$ using Eq.(\ref{eq.26}) we obtain: 
\begin{eqnarray}
- D(p_1+q+p_2-p'_2) & = & - D(q+p_{A}-p_{A-1}+p'_3-p_3) \nonumber \\ 
& = &  (q+p_{A}-p_{A-1}+p'_3-p_3)^2-m^2 + i\epsilon  \approx \nonumber \\
& &   2q\left[{q_0\over q}(E'_3-E_3)-(p'_{3z}-p_{3z})+i\epsilon\right]
= \left[(\Delta_{3} - (p'_{3z}-p_{3z}) + i\epsilon\right]. \nonumber \\
\label{eq.28}
\end{eqnarray}
In the derivation of Eq.(\ref{eq.28}) we use the kinematic condition
for the quasielastic scattering: $(q+p_A-p_{A-1})^2=m^2$ and define 
$\Delta_{3} = {q_0\over q}(E'_3-E_3)$. Similar to  the single rescattering case,
introducing $f_{NN}$ amplitudes one obtains: 
\begin{eqnarray}
A^\mu_{2a} & = & {F\over 4} 
\int \psi^+_{A-1}(p'_2,p'_3)
{f^{NN}(p'_3-p_3)\over \Delta_{3} - (p'_{3z}-p_{3z})+i\epsilon} 
{f^{NN}(p'_2-p_2)\over p^{m}_z+\Delta-p_{1z}+i\epsilon}
\times \nonumber  \\ 
&  & j^\mu(p_1,q)\psi_A(p_1,p_2,p_3)
{d^3p_1\over (2\pi)^3}{d^3p_3\over (2\pi)^3}{d^3p'_3\over (2\pi)^3}.
\label{eq.29}
\end{eqnarray}
To complete the calculation of double scattering amplitude one 
should calculate also the amplitude corresponding to the diagram of Figure 20b.
This amplitude can be obtained by interchanging momenta 
of nucleon ``2'' and ``3''. Doing so, for the complete double scattering amplitude 
one obtains:
\begin{eqnarray}
A^\mu_{2} & = & A^\mu_{2a}+A^\mu_{2b} = {F\over 4} \int \psi^+_{A-1}(p'_2,p'_3)
\times \nonumber \\
& & \left[{f^{NN}(p'_3-p_3)\over \Delta_{3} - (p'_{3z}-p_{3z})+i\epsilon} 
{f^{NN}(p'_2-p_2)\over p^{m}_z+\Delta-p_{1z}+i\epsilon} + 
{f^{NN}(p'_2-p_2)\over \Delta_{2} - (p'_{2z}-p_{2z})+i\epsilon}
{f^{NN}(p'_3-p_3)\over p^{m}_z+\Delta-p_{1z}+i\epsilon}\right]\nonumber \\
&  & j^\mu(p_1,q)\psi_A(p_1,p_2,p_3)
{d^3p_1\over (2\pi)^3}{d^3p_3\over (2\pi)^3}{d^3p'_3\over (2\pi)^3},
\label{F2}
\end{eqnarray}
were $\Delta_{2} = {q_0\over q}(E'_2-E_2)$.  Similar to the case of 
single rescattering, one can now generalize 
Eq.(\ref{F2}) for the case of $A>3$ which will read:
\begin{eqnarray}
A^\mu_{2}  & = &{F\over 4} \sum\limits_{i\ne j,=2}^A \int 
\psi^+_{A-1}(p_2,...p'_i,p'_j,...p_A)
 {f^{NN}(p'_j-p_j)\over \Delta_{j} - (p'_{jz}-p_{jz})+i\epsilon} 
{f^{NN}(p'_i-p_i)\over p^{m}_z+\Delta-p_{1z}+i\epsilon} \nonumber \\
& & j^\mu(p_1,q)\psi_A(p_1,...,p_i,p_j,...,p_A)
{d^3p_1\over (2\pi)^3}{d^3p_j\over (2\pi)^3}{d^3p'_j\over (2\pi)^3},
d^3p_2,...''...,d^3p_A,
\label{F2A}
\end{eqnarray}
where $''$  in the integration means that it does not contain $d^3p_id^3p_j$.

\subsection{Relation to the Glauber Theory}

It is interesting to see how the second order rescattering amplitude derived 
within GEA is related to the conventional Glauber theory. For this 
it is convenient to consider the double scattering amplitude in the 
coordinate space. Similar to the case of single rescattering we use coordinate 
space representation of nucleon propagators (as in Eq.(\ref{eq.10})) as well
as the wave function of ground and final states. 
Separating the c.m. and relative coordinates of the recoiled two nucleon system  
(see for details \cite{FSS97}) for Eq.(\ref{F2}) one obtains:
\begin{eqnarray}
\hat A^\mu_2 
& = & \int d^3x_1 d^3x_2 d^3x_3 \phi_A(x_1,x_2,x_3) 
F^{em}_1(Q^2) 
{\cal O}^{(2)}(z_1,z_2,z_3,\Delta_0,\Delta_{2},\Delta_{3})\nonumber \\
& & \Gamma^{NN}(x_2-x_1,\Delta_0)\Gamma^{NN}(x_3-x_1,\Delta_0)
e^{-i{3\over 2}\vec x_1 \cdot\vec p_{m}} \phi^{\dag}(x_2-x_3),
\label{eq.35}
\end{eqnarray}
where the profile functions $\Gamma_{NN}$ defined as in Eq.(\ref{eq.12}) and we 
introduce the ${\cal O}$ function which accounts for the geometry of two 
sequential rescatterings as:
\begin{eqnarray} 
{\cal O}^{(2)}(z_1,z_2,z_3,\Delta_0,\Delta_{2},\Delta_{3}) & = &  
\nonumber \\ 
& &  \Theta(z_2-z_1)\Theta(z_3-z_2)e^{-i\Delta_{3}(z_2-z_1)}
e^{i(\Delta_{3}-\Delta_0)(z_3-z_1)} \nonumber \\ 
& & + \Theta(z_3-z_1)\Theta(z_2-z_3)e^{-i\Delta_{2}(z_3-z_1)}
e^{i(\Delta_{2}-\Delta_0)(z_2-z_1)}.
\label{eq.36}
\end{eqnarray}
In the conventional Glauber theory ${\cal O}$ corresponds to the 
$\Theta(z_2-z_1)\Theta(z_3-z_1)$ which has a simple geometrical 
interpretation of sequential rescattering of knock-out nucleon 
on the two spectator nucleons in the target. It ensures that the rescattering 
can not happen with the nucleons which are located before the knocked-out nucleon. 
However the account for the finite momenta of target nucleons complicates the 
simple picture in the Glauber theory and the result is the Eq.(\ref{eq.36}). 
It is important to note that in the limit of the small Fermi momenta:
\begin{equation}
 {\cal O}|_{\Delta,\Delta_2,\Delta_3\rightarrow 0} 
\rightarrow \Theta(z_2-z_1)\Theta(z_3-z_1).
\end{equation}
To summarize the result of the calculation of double rescattering amplitude 
we like to point out that generalization of Glauber theory corresponds to 
the modification of profile functions according to Eq.(\ref{eq.12}) as well as 
generalization of the product of $\Theta$ functions to the ${\cal O}$ function of 
Eq.(\ref{eq.36}). 

\subsection{The Case of $A>3$ Nuclei}

Generalization of the scattering amplitude for the case of multiple 
rescatterings is straightforward\cite{FSS97}: 
\begin{eqnarray}
\hat T_{FSI}^{(n)} = \sum\limits_
{i,j, ..n=2;i\ne j\ne .. n}^{A} 
{\cal O}^{(n)}(z_1,z_i,z_j,...z_n,
\Delta_0,\Delta_{i},\Delta_{j}...\Delta_{n}) \times
\nonumber \\
\Gamma^{NN}(x_i-x_1,\Delta_0)\cdot\Gamma^{NN}(x_j-x_1,\Delta_0)\cdot...\cdot
\Gamma^{NN}(x_n-x_1,\Delta_0),
\label{eq.38}
\end{eqnarray}
where 
\begin{eqnarray}
{\cal O}^{(n)}(z_1,z_i,z_j,...,z_n\Delta_0,
\Delta_{i},\Delta_{j}...\Delta_{n}) = 
\sum\limits_{perm} \Theta(z_i-z_1)\Theta(z_j-z_i)...\Theta(z_n-z_{n-1})\times
\nonumber \\ 
e^{i(\Delta_0 - \Delta_j-...\Delta_n)(z_i-z_1)} 
e^{i\Delta_j(z_j-z_1)}...e^{i\Delta_n(z_n-z_1)}
e^{-i\Delta_0(z_i+z_j+...z_n - n\times z_1)}.
\label{eq.39}
\end{eqnarray}
The sum in Eq.(\ref{eq.39}) goes over all permutations between $i,j,...n$. 
We would like to draw attention that the contribution of diagrams where 
ejected nucleon interacts with say nucleon "2" then with nucleon "3" 
and then again with nucleon "2" is exactly zero. In a coordinate 
representation this follows  from the structure of the product of 
$\Theta$-functions. In the  momentum representation this follows from 
the possibility to close the contour of integration in the complex plane 
without encountering nucleon poles (see e.g. discussion in Section II.C of
Ref.\cite{FPSS} and the prove of the reduction theorem in Sec.VI.2).

It is easy to check that  in the case of small excitation energies i.e. 
($\Delta_0$, $\Delta_{i}$,$\Delta_{j}$ ... $\Delta_{n}\rightarrow 0$):
\begin{equation} 
 {\cal O}^{(n)}(z_1,z_i,z_j,..z_n,\Delta_0,\Delta_{i},\Delta_{j},...
\Delta_n)
\mid_{\Delta_0, \Delta_{i,k,n}\rightarrow 0 } 
\Rightarrow \Theta(z_i-z_1)\Theta(z_j-z_1)...\Theta(z_n-z_1),
\label{eq.40}
\end{equation}
and Eq.(\ref{eq.38}) is reduced  to the conventional form of 
the Glauber approximation, with a simple product of the $\Theta$-functions.
Within this particular approximation the sum over all $n$-fold rescattering
amplitudes can be represented in the form of an optical potential.

However, in many cases in high-energy $(e,e'N)$ reaction the excitation energies 
are not too small. The use of the 
${\cal O}^{(n)}(z_1,z_i,z_j,...z_n,\Delta_0,\Delta_{i},\Delta_{j},...\Delta_n)$, 
defined within GEA according to Eq.(\ref{eq.39}) instead of a simple product 
of $\Theta$ functions is the generalization of the nonrelativistic Glauber
approximation to the processes where comparatively large excitation
energies are important.

The practical consequence of the difference between ${\cal O}^{(n)}$
and the usual $\Theta$ functions is that for  sufficiently large 
excitation  energies the  sum of   $n$-fold rescatterings  
differs substantially from  the simple optical model limit. 
To illustrate the deviations from the conventional Glauber 
approximation (which is expressed by using a  simple product 
of the $\Theta$ functions) in Figure 21  
we compare ${\cal O}^{(2)}(z_1,z_2,z_3,\Delta_0,\Delta_{1},\Delta_{2})$ 
function with $\Theta(z_2-z_1)\Theta(z_3-z_1)$ for $(e,e'p)$
scattering  off $^3He$ target.
We use the kinematics for three body
breakup in the final state. Figure 21 demonstrates a considerable  
deviation between  ${\cal O}^{(2)}$ and the product of $\Theta$-functions
already at  comparatively low excitation energies.

\begin{figure}[h]  
\vspace{-0.2cm}  
\begin{center}  
\epsfig{angle=0,width=4.4in,file=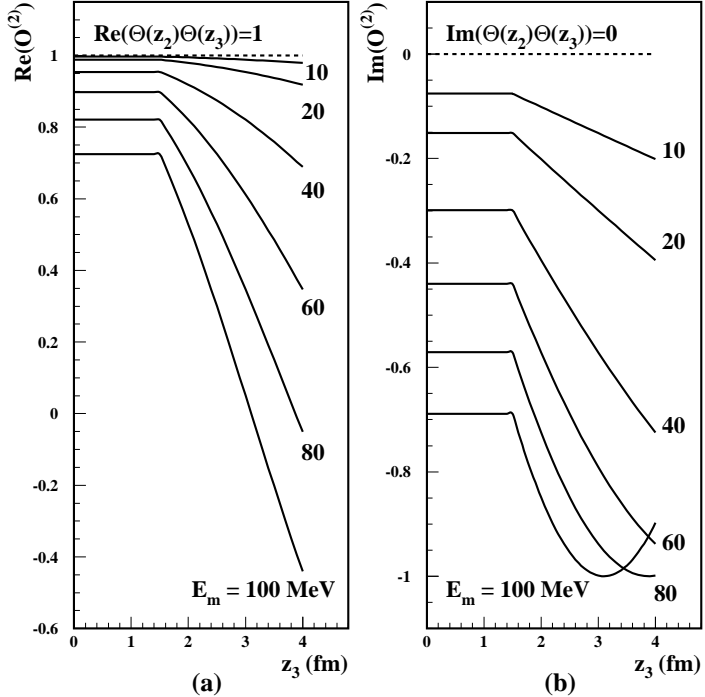}  
\end{center}  
\caption{Dependence of   ${\cal O}^{(2)}(z_1,z_2,z_3,\Delta_0,\Delta_{1},
\Delta_{2})$ and $\Theta(z_2-z_1)\Theta(z_3-z_1)$  on $z_3$ for  
different values of missing energy $E_m$ for $z_1=0$, $z_2=1.5~Fm$ 
and  $\Delta_{1}=\Delta_{2}=0$. 
a) Comparison of $Re {\cal O}^{(2)}(...)$(solid line) with  
$Re\Theta(z_2-z_1)\Theta(z_3-z_1)=1$(dashed line) and b) comparison  
of $Im {\cal O}^{(2)}(...)$(solid line) with 
$Im\Theta(z_2-z_1)\Theta(z_3-z_1)=0$ (dashed line).} 
\label{Fig.21}  
\end{figure}

 For example, the   
real parts differ by more than  $20\%$ already for $\sim 60 ~MeV$, 
leading to comparable difference of the double rescattering amplitude 
calculated including effects of longitudinal momentum transfer.
The detailed numerical studies of these effects for $A=3$ nuclei are 
in progress.

The comparisons in Figure 21 shows also that for light nuclei the conventional 
Glauber approximation, which neglects nuclear Fermi motion, is applicable only  
in the case of small values of excitation energies of the residual nuclei.

 \section{FSI and the study of short-range nucleon correlations in nuclei}

In this section we discuss how GEA allows us to gain an additional insight into the 
problem of the studies of short range correlations (SRC) in nuclei.

It is generally believed that experimental condition 
$|\vec p_m|=|\vec p_f  - \vec q| > k_{F}$, (where $k_{F}\sim 250~MeV/c$ 
is momentum of Fermi surface for a given nucleus) will enhance 
the contribution to the cross section from  the short-range 
nucleon correlations in the nuclear wave function. However 
simple impulse approximation relation (Eq.(\ref{IAm})) is, 
in general, distorted by the FSI, which, in average, is sensitive to the 
long range distances in nuclei. Let us denote the internal 
momentum of the  knock-out nucleons prior to the collision 
as $\vec p_1(p_{1z},p_{1t})$. It follows from Eqs.(\ref{F1m}),(\ref{F2A}):
\begin{equation}
\vec p_{1t} = \vec p_{mt} - \vec k_t,
\label{kt}
\end{equation}
where  $\vec p_{mt}$ the transverse component of the measured missing 
momentum, and $k_t$ is the  momentum transferred in rescattering.
The average $<\vec k_t^2 >\sim 0.25~ GeV^2$ in the integral over $k_t$; it is   
determined by the slope of the $NN$ amplitude. The longitudinal 
component  of the nucleon momentum in the initial state can be 
evaluated through  its value at the pole of the rescattered 
nucleon propagator (see e.g. Eqs.(\ref{F1_3},\ref{F1m}),(\ref{F2A})):
\begin{equation}
p_{1z}    =   p^z_m + \Delta_0
\label{eq.pole}
\end{equation}
where $p^z_m$ is longitudinal component of the  measured missing momentum  
and $\Delta_0$ is proportional to the excitation energy of the residual nuclear 
system (see Eqs.(\ref{Delta},\ref{Delta2},\ref{Delta3})) and is always positive.
Thus, if the measured  $p_{zm} >k_{F}$  then, $p_{1z}$ is even larger, i.e. 
($p_{z1} > p_{zm}$) and therefore the FSI amplitude is as sensitive to  
the short range correlations as the IA amplitude. 
In particular, within the approximation, where the high-momentum component 
of the nuclear spectral function is due to two-nucleon short-range 
correlations (see e.g. \cite{FS81,CFSM}), the condition $p_{zm}>k_{F}$ 
corresponds to projectile electron scattering  off the forward moving 
nucleon of the two-nucleon correlation accompanied by the emission of 
the backward nucleon.

The situation is the opposite  if the measured momentum $p_{zm} < - k_{F}$.
It follows from Eq.(\ref{eq.pole}), that in this case the momenta in 
the wave function contributing to the rescattering amplitude are smaller 
than those for IA: $|p^z_1| = |p^z_m|-\Delta_0  < |p^z_m|$. 

Experimentally, this situation corresponds to the forward nucleon 
electro-production at   ${Q^2\over 2mq_0}\equiv x>1$. An important new  
feature in this case is that in exclusive electro-production 
the value of $\Delta_0$ is measured  experimentally and can be easily chosen 
so  that  momenta entering in the  ground state wave function would be larger 
than  $k_{F}$. Therefore to investigate the short-range correlations in the  
$(e,e'p)$ reactions for   $x>1$, we have  to impose an additional condition: 
\begin{equation}
p^z_m -\Delta_0 > k_{F}
\label{addx}
\end{equation}
to suppress the contribution from large inter-nucleon  distances. The appearance 
of this constraint is the result of the GEA and can not be deduced within 
conventional Glauber approximation. 

There are several experimental projects  which will be able to check the 
prediction on the possibility of suppression of FSI at $x>1$ kinematics
\cite{KG,EGM,WKV,UJ,BBW,ESV,Kim,Eli,Templon}. 
The most basic experiments are the electrodisintegration of the deuteron 
\cite{KG,EGM,WKV,UJ} 
which will allow the measurement of the angular dependence of the transparency, 
T as it was defined in Eq.(\ref{T}) 
and check on the features discussed in the Sec.VIII.D, Figures 14 and 15.  
These experiments also will allow the  verification of the expectation that with 
increasing energies the qualitative changes discussed in Sections III-VI occur.

The experiments on high momentum transfer electro-disintegration of $A=3$ 
nuclei\cite{BBW,ESV} with large values of missing momenta and energy are 
important to verify the applicability of GEA for higher order of rescatterings. 
These measurements which are currently in the stage of data analyses,  will allow 
to identify the regions where FSI is suppressed thereby allowing 
to gain rather direct 
information about the short range structure of few-body nuclear systems

The possibility of suppression of FSI at $x<1$ and $x>1$ regions  will be explored 
in the experiments 
of Refs.\cite{Kim,Eli}, in which the attempt to extract the direct information about 
short-range few-nucleon correlations in semi-exclusive $A(e,e'NN)X$ reactions will be 
made.

Another interesting possibility is the study of the structure of FSI  
in high-energy two-body break-up processes in $^4He(e,e'p)^3H$ reactions. 
The momentum distribution for these reactions calculated within PWIA exhibits a 
minimum (node) at $p_m\approx 420 MeV/c$.

\begin{figure}[h]  
\vspace{-0.4cm}  
\begin{center}  
\epsfig{angle=0,width=3.2in,file=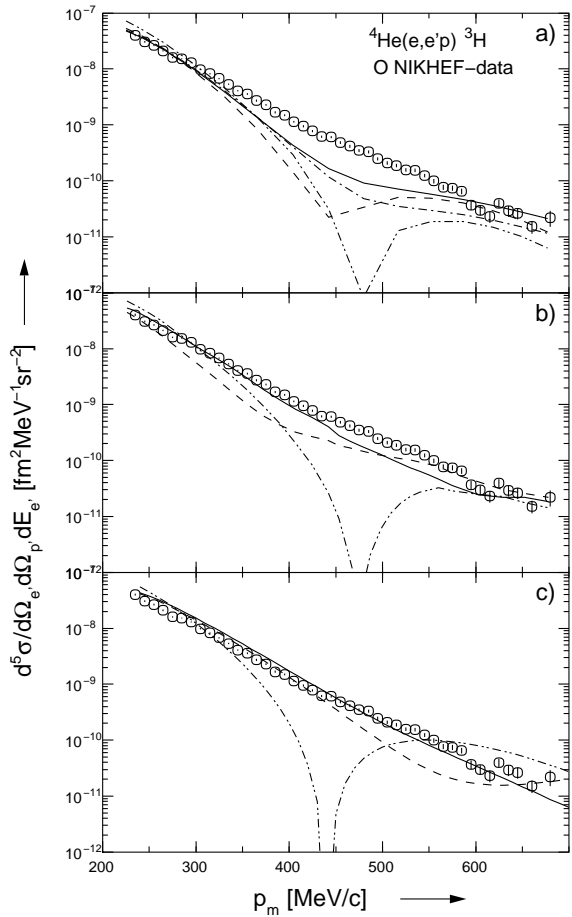}  
\end{center}  
\caption{Differential $^4He(e,e'p)^3H$ cross section as a function of the 
missing momentum $p_m$. The calculations, in (a) according to Ref.~[96]
%\cite{Laget} 
(dotted curve PWIA, dashed:+FSI, dot-dashed: +two-body MEC, solid: + tree-body MEC),
in (b) according to Ref.~[97] 
%\cite{Rocco} 
(dotted curve: PWIA, dashed +FSI, solid: + two-body 
currents) and in (c) according to Ref.~[98]
%\cite{Nagorny} 
(dotted curve: PWIA, dashed: tree-type 
diagrams, solid: + one-loop diagrams). The figure is from Ref.~[95].}
%\cite{vanLeeuwe}}
\label{Fig.22}  
\end{figure}

It is because the selection rules on angular momentum and parity requires only the  
S-wave to contribute in the two-body break-up amplitude of   $^4He(e,e'p)^3H$ 
reactions. 
The minimum in the $S$ wave  exists only if NN correlation are included into 
considerations. The lack of the minimum in the experimental distribution measured in 
Ref.(\cite{vanLeeuwe}) at initial electron energies $0.525~GeV$ correctly 
attributed to FSI, MEC and IC contributions (see Figure 22). 
If the minimum is indeed there then one expects that with an increase of 
energy these reactions, measured in parallel and perpendicular kinematics,  can be an 
important tool to study the structure of FSI. The naive estimation\cite{Templon} of the 
effects of FSI within GEA for  $^4He(e,e'p)^3H$ reactions at $Q^2= 1 GeV^2$ shows that 
one may expect qualitative differences between momentum distribution in parallel, 
antiparallel and perpendicular kinematics (Figure 23). 
Note that the calculations of 
Figure 23 are by no means to be considered complete since GEA is implemented using 
simple harmonic oscillator wave functions for the rescattering amplitudes. 
Also only the imaginary part of the NN scattering amplitudes is taken into account.
The realistic wave function in these calculations is used only for the PWIA 
contribution.

\begin{figure}[h]  
\vspace{0.2cm}  
\begin{center}  
\epsfig{angle=0,width=3.6in,file=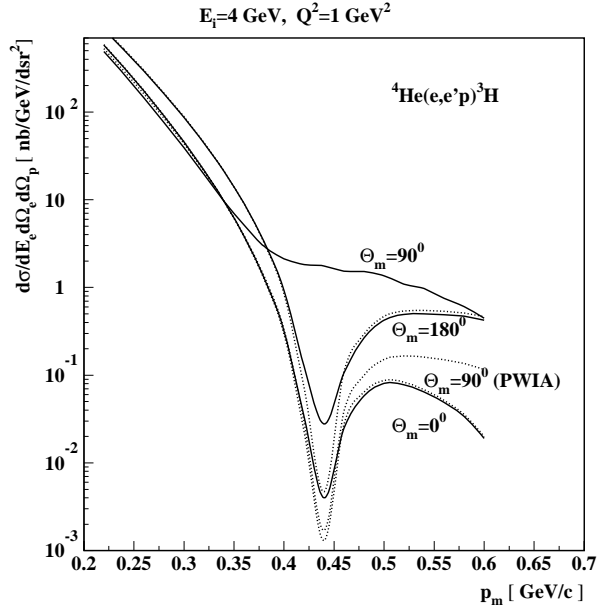}  
\end{center}  
\caption{Differential $^4He(e,e'p)^3H$ cross section as a function of the 
missing momentum $p_m$ at different values of missing momentum angles with respect to 
${\bf q}$. Dotted curves are PWIA prediction, solid curves PWIA+FSI.
Note that nonzero values of  minima in PWIA curves are the result  of 
the finite resolution of $p_m$ implemented in calculation. } 
\label{Fig.23}   
\end{figure}

Recent calculations of  $^4He(e,e'p)^3H$ reactions using realistic wave function in all 
rescattering amplitudes\cite{BNSUZ,ABCKMT} within a conventional Glauber approximation 
also demonstrate the large differences between the momentum dependence of the cross 
section at  different angles. However 
Ref.\cite{BNSUZ} predicts that some signature of the minimum still should 
be visible after taking into account for FSI while 
Ref.\cite{ABCKMT} predicts a complete disappearance of the minimum both for parallel and 
antiparallel kinematics. Thus the experimental data being analyzed in 
Ref.\cite{Templon} will be  of utmost importance.

\section{Emergence of Light Cone Dynamics}

The obtained within GEA results have a simple explanation in terms of the light-cone 
dynamics of  high-energy  scattering processes. Indeed, according to 
Eqs.(\ref{Delta},\ref{Delta2},\ref{Delta3}) $\Delta_0$  does not disappear  with 
an increase of energy. 
Hence the non-conservation of the longitudinal momentum of nucleons 
given by Eq.(\ref{eq.pole}) : $p_{1z}-p^z_m=\Delta_0$ remains finite in the 
high-energy limit. However, the rescattering of an energetic knock-out
nucleon practically does not change the "-" component of its four-momentum 
$p_- \equiv E -p_z$, (recall that $p_-$ is the longitudinal momentum as defined in 
light-cone variables, where $p^{\mu}\equiv (p_+,p_-,p_t)$ with 
$p_{\pm}=E\pm p_{z}$).
Really, if we define $p_{1-} = m - p_{1z}$ and $p_{m-} = p_{f-} - q_- = 
m - E_{m} - p_{mz}$, where $E_m = m + E_{A-1}-M_A$  is the missing energy, 
then according to Eq.(\ref{eq.pole}) the non-conservation of "-" component is:
\begin{equation}
p_{1-}-p_{m-} \approx {Q^2\over 2 q^2}E_m = {E_m\over 2(1+{q_0\over 2mx})}. 
\label{nonconv}
\end{equation}
It vanishes with increase of the  virtual photon energy $q_0$. Hence, 
the  physical interpretation of Eq.(\ref{eq.pole}) is that at high energies 
elastic FSI does not change noticeably the  light-cone "-" component of 
the struck  nucleon momentum.   This result is in agreement with our observation
in Sec.VI.1.
This reasoning indicates that description  
of the  FSI in high-energy processes should be simplified when treated
within the framework of the light-cone dynamics. Our previous analysis
of  $x>1$, large $Q^2$ data on inclusive $(e,e')$ processes 
is consistent with
this idea\cite{Day}. 

The above observation helps to rewrite the deduced formulae in the form  
accounting for, in a straightforward way, the fact that high-energy processes 
develop along the light-cone.  
One uses the previously introduced light-cone momenta 
$\alpha_i\equiv A{p_{i-}\over P_{A-}}$. 
Here $\alpha_i/A$ is a momentum fraction of target 
nucleus carried by the nucleon-$i$. Using the above discussed expressions for 
$p_{m-}$ and $p_{1-}$ and Eqs.(\ref{F1m},\ref{Delta2},\ref{Delta3}) for the propagator 
of a  fast nucleon we obtain: 
\begin{equation}
{1\over [p^{m}_z + \Delta_0 
- p_{1z} + i\epsilon]} = {1\over m[\alpha_1 - \alpha_m +{q_0-q\over qm}E_m
+ i\epsilon]} \approx {1\over m[\alpha_1 - \alpha_m -{Q^2\over 2q^2}{E_m\over 
m}+ i\epsilon]}.
\label{lcprop}
\end{equation}
In the kinematics where relativistic effects in the wave function of the 
target and residual nucleus are small and $\alpha_j \approx 1-{p_{jz}\over m}$,  
there is a smooth correspondence between nonrelativistic and light-cone 
wave functions of the nucleus\cite{FS81}, 
i.e. $\phi_A(p_1,...p_j,..p_A)\approx \phi_A(\alpha_1,p_{1t}, ... 
\alpha_j,p_{jt},... \alpha_A,p_A)/m^{A\over 2}$. Therefore  the amplitude of single 
rescattering - Eq.(\ref{eq.12}) can be rewritten as:
\begin{eqnarray} 
T^{(b)} = -{\sqrt{(2\pi)^3}(2\pi)^3 \over  2m } 
\int \psi_A(\alpha_1,p_{1t},\alpha_2,
p_{2t},\alpha_3,p_{3t})F^{em}_1(Q^2){f^{NN}\over 
[\alpha_1 - \alpha_m -{Q^2\over 2q^2}{E_m\over m}
+ i\epsilon]}\nonumber \\
 \psi_{A-1}(\alpha'_{2},p'_{2t},\alpha_3,p_{3t}) 
{d\alpha_1 d^2p_{1t}\over (2\pi)^3} {d\alpha_3 d^2p_{3t}\over (2\pi)^3}.
\label{eq.12lc}
\end{eqnarray}
where according to Eq.(\ref{eq.26}) $\alpha_2 = \alpha'_2 = 3 -\alpha_1-\alpha_3$.  
Eq.(\ref{eq.12lc}) shows that in the limit when ${Q^2\over 2q^2}
{E_m\over m}\rightarrow 0$, the amplitude $T^{(b)}$ is expressed through  
the light-cone variables and the light-cone wave functions of nucleus. 
Note that   the eikonal scattering corresponds to the linear (in $\alpha_1$) 
propagator of the fast nucleon. It is instructive that the regime of the 
light-cone  dynamics is reached in Eq.(\ref{eq.12lc}) at relatively moderate 
energies. Indeed, let us consider kinematics when $\alpha_1$ is close to unity, 
(which is the  case in our analysis). At $q_0\sim 2~GeV$, 
${Q^2\over 2q^2}{E_m\over m}= {1\over 2(1+{q_0\over 2mx})}{E_m\over m} 
\sim (0.05-0.07)\ll 1$. For estimate we take $x=1$ and for missing energy 
$E_m\sim 0.2-0.3~GeV$ which is close to the limit of applicability of 
the description of nuclei as a many-nucleon system (e.g. \cite{FS88}).   
Similar reasoning is applicable for the double rescattering 
amplitude in Eq.(\ref{eq.29}). Here we obtain:
\begin{eqnarray}
T^{(d)} & = & {\sqrt{(2\pi)^3}(2\pi)^3\over 4 m^2} 
\int
\psi_A(\alpha_1,p_{1t},\alpha_2,p_{2t},\alpha_3,p_{3t})F^{em}_1(Q^2)
\times \nonumber  \\ 
&  & {f^{NN}(p_{1t}-p_{mt}-(p'_{3t}-p_{3t}))\over 
[\alpha_1 - \alpha_m -{Q^2\over 2q^2}{E_m\over m} + i\epsilon]}
 {f^{NN}(p'_{3t}-p_{3t})\over 
[\alpha_3 - \alpha'_3 -{Q^2\over 2q^2}{k^2_{3t}\over2 m^2} + i\epsilon]}
\psi_{A-1}(\alpha_2,p'_{2t},\alpha_3,p'_{3t})
\nonumber \\
& & {d\alpha d^2p_{1t}\over (2\pi)^3}{d\alpha_3 d^2 p_{3t}\over (2\pi)^3}
{d\alpha'_3 d^2p'_{3t}\over (2\pi)^3}.
\label{eq.29lc}
\end{eqnarray}

Another interesting consequence of the  representation of the scattering 
amplitude through the light-cone variables, is the simple form of the
closure approximation for the  sum over the residual $(A-1)$ nuclear 
states in the $A(e,e'N)(A-1)$ reaction. 
When summing over the $E_m$ at fixed $p_m$ the rescattering amplitudes 
(eg. Eq.(\ref{eq.12})) could not be factored out from the sum because they 
depend on $E_m$ through the $\Delta$ factors (Eqs.(\ref{Delta},\ref{Delta2},
\ref{Delta3})).
In the case of the light-cone representation (e.g. Eq.(\ref{eq.12lc})) the 
analogous  procedure\cite{FS88} is to sum over   $p_+\approx m+E_m+p_{mz}$ 
at fixed $\alpha_m$. It  follows from  Eq.(\ref{eq.12lc}) that in such 
a sum the scattering amplitude is independent of  $p_+$  and therefore the 
application of closure has a simple form.

Note that the present discussion of the  light-cone dynamics is 
by no means complete, since we don't consider the relativistic effects 
in description of the nuclear wave functions. The extension of the  
current analysis to the light-cone formalism will be presented 
elsewhere.

\section{Conclusions}

The increase of transferred energies provides a qualitative 
new regime in electro- nuclear reactions. The possibility to 
suppress the long-range  phenomena in these reactions opens a 
completely new window in the study of microscopic properties 
of nuclear matter at small distances.

We give arguments why with the increase of transferred energies the 
description of semi-exclusive electro-nuclear reactions 
should simplify.
However, in this regime  the theoretical 
methods of low/intermediate energy electro-nuclear reactions 
become inapplicable. 
The theoretical approach which allows us to do consistent calculation 
is based on the fact that in the high energy regime new small 
parameters (such as ${q_-\over q_+}$) emerge (Section II). 
We demonstrate how the existence of 
these parameters reveals several new features in electro nuclear reactions 
such as: the on-shellness of the ``good'' component of the electromagnetic current 
of bound nucleon, onset of a new approximate conservation law for high energy small 
angle rescatterings and the existence of a reduction theorem which allows one to 
group potentially infinite number of rescattering amplitudes into a finite number 
covariant amplitudes. 

Furthermore we identify the set of effective Feynman diagram rules which 
allow one to calculate these covariant amplitudes. This framework we call 
generalized eikonal approximation (GEA).

GEA provides a natural framework for generalization of the  conventional 
Glauber approximation to high-energy processes with large values of 
missing momenta and energy in the reaction.
This approach  adequately describes   the relativistic  dynamics 
characteristic to high-energy reactions.

To demonstrate the application of GEA we first calculate the 
high energy electrodisintegration of the deuteron.  Furthermore 
we discuss the electrodisintegration of $A=3$ targets and generalize 
the obtained formulae for the case of large $A$ nuclei.
It  follows from these considerations that 
the formulae of the conventional Glauber approximation are a  limiting
case of GEA  at small values of missing momenta and energy.

Analyzing obtained formulae for final state reinteractions 
we identify the  kinematic domains preferable 
for  investigation of short-range nucleon correlations in nuclei.
We found  a new  kinematic condition:
$p^z_m -\Delta_0 > k_{F}$, for semi-exclusive reactions to enhance the 
contribution of  short-range nucleon correlations at $x>1$
and reduce the FSI.

We also observe that dominance of  light-cone dynamics follows 
directly from the analysis of the Feynman diagrams, and that  the "-" 
component of the target nucleon momentum is almost conserved in FSI.
Therefore, by measuring the "-" component of missing momenta we 
directly tag the preexisting momenta in the light-cone nuclear wave 
function. At the end we demonstrate how simple are the rescattering 
amplitudes expressed through the  light-cone  momenta and most importantly 
the missing energy  (defined in the light-cone) dependence is factorized 
from the rescattering amplitudes. This observation has a significant consequence 
for application of the closure approximation in the discussion of FSI in high energy 
inclusive $(e,e')$ reactions.

Finally we reviewed briefly the ongoing and planned experiments on high-energy 
semi-exclusive reactions. Emerging experimental data on these reactions within the next
several years will significantly contribute to the  understanding of the dynamics 
of high-energy electro-nuclear reactions as well as provide unprecedented 
quality of data for investigation of nuclear structure at small internucleon distances.

\begin{acknowledgements}
I  would like to thank L.L.~Frankfurt, G.~A.~Miller  and M.I.~Strikman 
for  fruitful collaboration and for useful discussions and suggestions during 
the period of writing this review. Special thanks to E.~Henley for the suggestion 
to write this review and for very valuable comments on the text.
This work is supported  by  DOE grant under contract DE-FG02-01ER-41172.

I gratefully acknowledge also a contract from Jefferson Lab under which this work was 
done. The Thomas Jefferson National Accelerator Facility (Jefferson Lab) is operated 
by the Southeastern Universities Research Association (SURA) under DOE contract 
DE-AC05-84ER40150.
\end{acknowledgements}

\appendix

\section{Why closure approximation has wider region of applicability
in the Light-Cone as compared to the nucleus rest frame} 
In the calculation of $n$-fold rescattering amplitude of Figure 12 
we assumed the decoupling (from the excitation energies of 
intermediate states) of the propagator of high energy  knocked-out   
nucleon - $D(p_1+q)^{-1}$. Such a decoupling allows to  use the closure
over the sum over the excitations of intermediate nuclear states. 
As a result the scattering  amplitude in Eq.(\ref{amp_n}), is calculated 
in terms of the propagators of free spectator nucleons in the 
intermediate states.

To visualize the conditions when the decoupling of high-energy 
part of the diagram of Figure 12 from low-energy part would be valid 
we consider two reference frame descriptions: Nucleus rest frame 
(Lab frame) and Light Cone.

In the Lab frame  the inverse propagator of energetic knocked-out  
nucleon: $-D(p_1+q)= (p_1+q)^2-m^2+i\epsilon$ can be written as:
\begin{equation}
(p_1+q)^2-m^2 = p_1^2 + 2E_1q_0 - 2\vec p_{1}\cdot{\vec q} + q^2 - m^2 = 
2|\vec q|\left[{p_1^2-m^2\over 2|\vec q|} + E_1{q_0\over |\vec q|} -
    p_{1z} - {Q^2\over |\vec 2q|}\right]
\label{a1}
\end{equation}
It follows from the right hand side of Eq.(\ref{a1}) that only the term 
$E_1{q_0\over |\vec q|}- p_{1z}$ survives in the limit of large momentum transfer 
($\vec q$) and fixed $x_{Bj}$. Thus in high energy  limit within Lab frame 
description one should retain the dependence of propagator on the 
excitation energy of intermediate state (via ${E_1}$). Therefore,
unless  the $E_1$ dependence of the propagator of knocked-out nucleon 
can be neglected the use of closure over the intermediate nuclear 
states can not be justified. In the Lab frame description such a 
neglection is legitimate in the nonrelativistic limit only where the 
term  ${p_1^2\over 2m^2}\ll 1$ is neglected everywhere in the expression 
of the scattering amplitude.
Such a restriction on the applicability of the closure for the sum over
the intermediate states is of crucial importance for the  models where 
relativistic effects are treated on the basis of the Lab frame description.

Above calculation does not take into account  additional approximate
conservation law characteristic for light-cone dynamics.
Let us introduce  light-cone momenta for  four-vectors as: 
$p^\mu(p_+,p_-,p_t)$, where $p_{\pm} = E\pm p_{z}$. Using these definitions, 
for the inverse propagator of knocked-out nucleon  one obtains the form:
\begin{equation}
(p_1+q)^2-m^2 = p_1^2 + p_{1+}q_{-} + p_{1-}q_{+} +q^2 - m^2 = 
q_{+}\left[{p_1^2-m^2\over q_+} + p_{1+}{q_{-}\over q_{+}} + 
p_{1-} - {Q^2\over q_{+}}\right].
\label{a2}
\end{equation}
As follows from the above equation the only term, that survives 
at fixed $x_{Bj}$ and high energy transfer limit, is $p_{1-}$. 
Therefore at fixed $p_{-}$ we found  effective factorization of  
high-energy propagator from low energy intermediate nuclear part 
whose excitation energy on light cone is defined  by the 
$p_{1+}$ \cite{FS81,FS88}.  Such a decoupling  applies for any 
values of Fermi momenta of  the target 
nucleon (no restriction like ${p_1^2\over 2m^2}\ll 1$ is formally needed). 
Therefore it is possible  to extend the application of the 
closure  over intermediate  states of the residual nucleus to 
the domain of relativistic  momenta of target nucleons.  The price 
is to introduce  the light-cone wave function's of the target (similar 
to the case of pQCD).

Note that the considerations in  present work are restricted by relatively 
small Fermi momenta (e.g. Eq.(\ref{kind})) since  we use  ${p_1^2\over 2m^2}\ll 1$
in the scattering amplitude. For larger Fermi momenta a way to generalize 
the obtained results is to use light-cone description, which
is out of scope of the present paper. Note that light-cone mechanics 
of nuclei is rather similar to the nonrelativistic ones\cite{FS81,FS91}.

\section {Analytic calculation or rescattering amplitude}
\label{B}

We   calculate the rescattering amplitude 
in Eq.(\ref{F1_3}) by the method described in  Ref.\cite{FGMSS95} 
using the deuteron wave function in  momentum space,  defined as\cite{BJ}:
\begin{equation}
\psi^\mu_D(p) = {1\over \sqrt{4\pi}}\left(u(p) + w(p)
\sqrt{1\over 8}S(p_z,p_t)\right)\chi^\mu,
\label{eq.b13}
\end{equation}
where $\chi^\mu$ is the deuteron spin function and 
\begin{equation}
S(p_z,p_t) = {3(\vec \sigma_p \cdot \vec p)(\vec\sigma_n 
\cdot\vec p)\over p^2} -  
\vec\sigma_p\cdot\vec\sigma_n,
\label{eq.b14}
\end{equation}
where $\sigma_p$, $\sigma_p$ Pauli matrices.
The functions $u(p)$ and $w(p)$ are the radial wave functions 
of $S$- and $D$- states, respectively and they can be  written
as\cite{paris,bonn}:
\begin{equation}
u(p) = \sum\limits_{j} {c_j\over p^2 + m^2_j}; \ \ \ \ \ \ \ \ 
w(p) = \sum\limits_{j} {d_j\over p^2 + m^2_j} 
\label{eq.b15}
\end{equation}
where $\sum\limits_j c_j = \sum\limits_j d_j = 0$, which guarantees that
$u(p),w(p)\sim {1\over p^4}$ at large 
$p$  and $\sum\limits_j {d_j\over m^2_j}=0$ to provide $w(p=0)=0$. 
Insertion of  Eq.(\ref{eq.b15}) 
into the Eq.(\ref{F1_3}) gives: 
\begin{eqnarray}
& & A_{1}^\mu  =   -{(2\pi)^{{3\over 2}\sqrt{2E_s}}\over 2}
\int {d^2p'_{st}\over (2\pi)^2} f^{pn}(k_t)\cdot j^{\mu}_{\gamma^*N}
\nonumber \\
& & \times\int {d p'_{sz}\over (2\pi)}
\left({c_j\over p'^2_s + m^2_j} + {d_j\over p'^2_s + m^2_j}\sqrt{1\over 8}
S(p'_{sz},p'_{st})\right)
{\chi^\mu  \over  p'_{sz}- p_{sz} + \Delta  + i\epsilon }. 
\label{eq.b16}
\end{eqnarray}  
Substituting    
$p'^2_s + m^2_j = (p'_{sz} + i\sqrt{m_j^2+p'^2_{st}})(p'_{sz} - 
i\sqrt{m_j^2+p'^2_{st}})$
one can perform the integration over $p'_{sz}$ by closing the 
contour in the upper 
$p'_{sz}$ complex semi-plane.
Note that the $p^{-2}$ dependence of the tensor function $S(p)$ will not 
introduce a new 
singularity, since $w(p=0)=0$. Setting the 
residue at the point $p'_{sz} = i\sqrt{m_j^2+p'^2_{st}}$ 
we obtain:
\begin{eqnarray}
& &  A_{1}^\mu  =   -{i(2\pi)^{{3\over 2}}\sqrt{2E_s}\over 2}
\int {d^2p'_{st}\over (2\pi)^2} f^{pn}(k_t) \cdot j^{\mu}_{\gamma^*N} 
\left[ {c_j\over 2i\sqrt{p'^2_{st} + m^2_j}} + \right. \nonumber \\
& &  \left. + {d_j\over 2i\sqrt{p'^2_{st} + m^2_j}}\sqrt{{1\over 8}} 
S(i\sqrt{p'^2_{st} + m^2_j},p'_{st})\right]
{\chi^\mu  \over  i\sqrt{p'^2_{st} + m^2_j}- p_{sz} + \Delta}. 
\label{eq.b17}
\end{eqnarray}
After regrouping  of the real and imaginary parts, the  above equation can be 
rewritten as:
\begin{equation}
 A_{1}^\mu  =   -{(2\pi)^{{3\over 2}}\sqrt{2E_s}\over 4i  } 
\int {d^2k_t\over (2\pi)^2} f^{pn}(k_t) \cdot j^{\mu}_{\gamma^*N} 
\left(\psi^\mu(\tilde p_s) - i \psi'^\mu(\tilde p_s)\right),
\label{eq.b18}
\end{equation}
where $\tilde p_s(\tilde p_{sz}, \tilde p_{s\perp})\equiv 
\vec { \tilde p_s}(p_{sz}-\Delta, 
\vec p_{st}-\vec k_t)$, $\psi^\mu$ is the wave function defined in 
Eq.(\ref{eq.b13}) and $\psi'^\mu$ is defined as:
\begin{equation}
\psi'^\mu(p) = \left(u_1(p)p_z + {w_1(p)p_z\over \sqrt{8}}S(p_z,p_t) + 
{w_2(p)\over \sqrt{8} p_z}\left[S(p_z,p_t) - S(0,p_t)\right]\right)\chi^\mu,
\label{eq.b19}
\end{equation}
where 
\begin{eqnarray}
u_1(p) &  = & \sum\limits_j{c_i\over \sqrt{p^2_t+m^2_j}(p^2+m^2_j)},
\ \ 
w_1(p) = \sum\limits_j{d_i\over \sqrt{p^2_t+m^2_j}(p^2+m^2_j)}, \nonumber \\ 
w_2(p) &  = & \sum\limits_j{d_i\over \sqrt{p^2_t+m^2_j}m^2_j}.
\label{eq.b20}
\end{eqnarray}
Note that the last term in Eq.(\ref{eq.b19}) does not have a
singularity at $p_z=0$ since $(S(p_z,p_t) - S(0,p_t))\sim p_z$.


\begin{references}

%\vspace{-0.2cm}

\bibitem{Feynman}R.P. Feynman, {\em Photon Hadron Interactions}, Benjamin Inc. 1972  
\bibitem{CW}H.~Cheng and T.~T.~Wu, ``Expanding Protons: Scattering At High-Energies,''
            {\it  Cambridge, USA: MIT-PR. (1987) 285p}.
%%%%%%%%% Proposals and experiments on high energy semi-exclusive reactions %%%%%%%%
\bibitem{PACS}''Program Advisory Committee'' at Jefferson Lab, Newport News, VA\\
                {\em http://www.jlab.org/exp\_prog/PACpage/index.html}.
\bibitem{Bochna}C.~Bochna {\it et al.}  [E89-012 Collaboration], 
                       Phys.\ Rev.\ Lett.\  {\bf 81}, 4576 (1998).
\bibitem{Bulten} H.~J.~Bulten {\it et al.}, Phys.\ Rev.\ Lett.\  {\bf 74}, 4775 (1995).
\bibitem{NE18}N.~Makins {\it et al.}, [NE18 collaboration], 
             Phys.\ Rev.\ Lett.\  {\bf 72}, 1986 (1994).
\bibitem{O'Neill}T.~G.~O'Neill {\it et al.},  [NE18 collaboration],  
                 Phys.\ Lett.\ B {\bf 351}, 87 (1995).
\bibitem{Abbot}D.~Abbott {\it et al.} Phys. \ Rev. \ Lett. {\bf 80}, 5072 (1998).
\bibitem{Ent} K.~Garrow {\it et al.}, arXiv:hep-ex/0109027, 2001.

\bibitem{Liyanage}N.~Liyanage {\it et al.}  [Jefferson Lab Hall A Collaboration],
                  Phys.\ Rev.\ Lett.\  {\bf 86}, 5670 (2001)
\bibitem{Airapetian}A.~Airapetian {\it et al.}  [HERMES Collaboration],
                    Eur.\ Phys.\ J.\ C {\bf 20}, 479 (2001).
\bibitem{Steenhoven}G.~van der Steenhoven  [HERMES Collaboration],
                   Nucl.\ Phys.\ A {\bf 663}, 320 (2000).
\bibitem{Ackerstaff} K.~Ackerstaff {\it et al.}  [HERMES Collaboration],
                     Phys.\ Rev.\ Lett.\  {\bf 82}, 3025 (1999)
%%%%%%%%%%%%%%%%%%%%%%%%%%%%%%%%%%%%%%%%%%%%%%%%%%%%%%%%%%%%%%%%%%%%%%%%%%%
\bibitem{wp}White Paper: "The Science Driving the 12 GeV Upgrade of CEBAF", 
            Jefferson Lab, Newport News, VA, 2000.
\bibitem{DR} M.~Duren and K.~Rith, Phys.\ Bl.\  {\bf 56N10}, 41 (2000).
\bibitem{OV} C.~D.~Epp and T.~A.~Griffy, Phys. Rev. {\bf C1}, 1633, (1970); 
             F.~Cannata, J.~P.~Dedonder and F.~Lenz, Ann. Phys. (N.Y.) 
             {\bf 143}, 84 (1982).
\bibitem{CEBAF}CEBAF Conceptual design report, Southeastern Universities Research 
               Association, Newport News, 1990.
\bibitem{HERMES} Technical Report of the HERMES Experiment, DESY, 1991 (unpublished).

\bibitem{Glauber} R.~J.~Glauber Phys.Rev. {\bf 100}, 242 (1955);
                  Lectures in Theoretical Physics, {\bf v.1}, ed. 
                  W.~Brittain and L.~G.~Dunham, Interscience Publ., 
                  N.Y. 1959.
\bibitem{Yennie} D.~R.~Yennie, in { Hadronic interactions of Electrons 
                 and photons}, edited by J.~Cummings and D.~Osborn 
                 (Academic, New York, 1971), p.321.
\bibitem{Moniz} E.~J.~Moniz and G.~D.~Nixon, Ann. Phys. (N.Y.) {\bf 67}, 58 
                (1971).
\bibitem{ABV}G.~D.~Alkhazov, S.~L.~Belostotsky and A.~A.~Vorobev, 
             Phys.\ Rept.\  {\bf 42}, 89 (1978).

\bibitem{F1} G.~R.~Farrar, L.~L.~Frankfurt, H.~Liu and M.~I.~Strikman 
             Phys. Rev. Lett. {\bf 61}, 686 (1988).
\bibitem{M1}  T.-S.~H.~Lee and G.~A.~Miller,  Phys.Rev. C {\bf 45}, 1863 (1992).
\bibitem{M2}  B.~K~Jennings and G.~A.~Miller,  Phys. Lett. {\bf B318}, 7 (1993).
\bibitem{Si}  A.~Kohama, K.~Yazaki, and R.~Seki, Nucl.Phys. {\bf A551}, 687 (1993).
\bibitem{Zv}  L.~L.~Frankfurt,  M.~I.~Strikman and M.~Zhalov, 
                  Phys. Rev. C {\bf 50}, 2189 (1994).
\bibitem{Br} O.~Benhar {\em et al.}, Phys. Rev. Lett. {\bf 69}, 881 (1992).
\bibitem{R2}  A.~S.~Rinat and M.~F.~Taragin,  Phys.Rev. {\bf C52}, 28  (1995).
\bibitem{N1} N.~N.~Nikolaev, A.~Szczurek, {\em et al.},  Phys. Lett. 
                {\bf B317}, 287 (1993).
\bibitem{Wt}  S.~Frankel, W.~Frati and N.~R.~Walet, Nucl. Phys. {\bf A580}, 
              595 (1994).
\bibitem{Bi}  A.~Bianconi, S.~Boffi, D.~E.~Kharzeev Phys. Lett. {\bf B325}, 
                294, (1994).
\bibitem{MP} E.~J.~Moniz, Summary Talk in PANIC-XIII, Perugia, Italy, 1993.
\bibitem{Ciofi1}C.~Ciofi degli Atti, L.~P.~Kaptari and D.~Treleani,
                Phys.\ Rev.\ C {\bf 63}, 044601 (2001)
\bibitem{Ciofi2}H.~Morita, C.~Ciofi degli Atti and D.~Treleani, 
                Phys.\ Rev.\ C {\bf 60}, 034603 (1999).
\bibitem{SJ}S.~Jeschonnek, Phys.\ Rev.\ C {\bf 63}, 034609 (2001)
\bibitem{FSZ}  L.~L.~Frankfurt,  M.~I.~Strikman and M.~Zhalov, 
                  Nucl. Phys.  {\bf A515}, 599 (1990).
\bibitem{FMSS95} L.~L.~Frankfurt,  E.~J.~Moniz, M.~M.~Sargsyan and 
                  M.~I.~Strikman, Phys. Rev. {\bf C51}, 3435 (1995).
\bibitem{FMS94}L.~L.~Frankfurt, G.~A.~Miller and M.~Strikman,
               Ann.\ Rev.\ Nucl.\ Part.\ Sci.\  {\bf 44}, 501 (1994).
\bibitem{Gribovi}V.N.~Gribov, JETP {\bf 29}, 483 (1969).
\bibitem{Saclay}C.~Marchand {\it et al.}, Phys.\ Rev.\ Lett.\  {\bf 60}, 1703 (1988);
                A.~Bussiere {\it et al.},Nucl.\ Phys.\ A {\bf 365}, 349 (1981).
\bibitem{Mainz} K.~I.~Blomqvist {\it et al.},Phys.\ Lett.\ B {\bf 424}, 33 (1998).
\bibitem{NIKHEF}  D.~L.~Groep {\it et al.}, Phys.\ Rev.\ C {\bf 63}, 014005 (2001).
\bibitem{Arenhovel}H.~Arenhovel, W.Leidemann and E.~L.~Tomusiak,
                   Phys.\ Rev.\ C {\bf 52}, 1232 (1995).
\bibitem{PP}V.R.~Pandharipande and S.C.~Pieper, Phys.\ Rev.\ C {\bf 45}, 791 (1992).  
\bibitem{LSFSZ} L.~Lapikas, G.~van der Steenhoven, L.~Frankfurt, M.~Strikman and 
        M.~Zhalov, Phys.\ Rev.\ C {\bf 61}, 064325 (2000).
\bibitem{FS81} L.~L.~Frankfurt, M.~I.~Strikman, Phys. Rep. {\bf 76} 215 (1981).
\bibitem{SG} Y.~Surya and F.~Gross, Phys.\ Rev.\ C {\bf 53}, 2422 (1996).
\bibitem{GMD}R.~Dashen and M.~Gell-Mann, Phys. Rev. Lett {\bf 17}, 340 (1966).
\bibitem{Gribov}V.~N.~Gribov, JETP, {\bf 30}, 709 (1970).
\bibitem{Grossi}F.~Gross, in {\em Modern Topics in Electron Scattering}, 
                Editors  B.~Frois and I.~Sick, World Scientific, 1991, p.219.
\bibitem{KochPol}H.~W.~Naus, S.~J.~Pollock, J.~H.~Koch and U.~Oelfke,
                 Nucl.\ Phys.\ A {\bf 509}, 717 (1990).
\bibitem{deFor} T.~De Forest, Nucl.\ Phys.\ A {\bf 392} (1983) 232.
\bibitem{Thomas} J.~Speth and A.~W.~Thomas, Adv.\ Nucl.\ Phys.\  {\bf 24}, 83 (1997).
\bibitem{Frodyma}M.~Frodyma {\it et al.}, Phys.\ Rev.\ C {\bf 47}, 1599 (1993).
\bibitem{FS88}L.L.~Frankfurt, M.I.~Strikman, Phys. Rep.  {\bf 160}, 235 (1988). 
\bibitem{Stoler} P.~Stoler, Phys.\ Rept.\  {\bf 226}, 103 (1993).
\bibitem{Azimov} Y.~I.~Azimov, Yad.\ Fiz.\  {\bf 11}, 206 (1970).
\bibitem{PDBC} P.D.B.~Collins, {An Introduction to Regge Theory and High 
               Energy Physics}, Cambridge University Press, Cambridge, 1977.
\bibitem{PDG} D.~E.~Groom {\it et al.}  [Particle Data Group Collaboration],
             Eur.\ Phys.\ J.\ C {\bf 15}, 1 (2000).
\bibitem{FGMSS95}L.~L.~Frankfurt, W.~R.~Greenberg, G.~A.~Miller, 
                 M.~M.~Sargsian and M.~I.~Strikman,
                 Z.\ Phys.\ A {\bf 352}, 97 (1995).
\bibitem{FSS97}L.~L.~Frankfurt, M.~M.~Sargsian and M.~I.~Strikman, 
               Phys.\ Rev.\ C {\bf 56}, 1124 (1997).
\bibitem{BC} L.~Bertuchi and A.~Capella   Il Nuovo Cimento, {\bf A51}, 369 (1967). 
\bibitem{Bert} L.~Bertuchi, Il Nuovo Cimento, {\bf A11},45 (1972). 
\bibitem{FSP} L.L.~Frankfurt and M.I.~Strikman, {\em Private Communication}.
\bibitem{KG}S.~E.~Kuhn and K.~A.~Griffioen (spokespersons), 
            {\em Electron Scattering from a 
            High Momentum Nucleon in Deuterium},Jefferson Lab Proposal- E-94-102.
\bibitem{EGM}K.~Sh.~Egiyan,  K.~A.~Griffioen and M.~I.~Strikman  (spokespersons), 
            {\em Measuring Nuclear Transparency in Double Rescattering Processes},
            Jefferson Lab Proposal -  E-94-019.
\bibitem{WKV}W.~Boeglin, A.~Klein and E.~Voutier (spokespersons), 
             {\em A Study of the Dynamics of 
             the Exclusive Electro-Disintegration of the Deuteron},
             Jefferson Lab Proposal -  PR-01-008.
\bibitem{UJ}P.~Ulmer and M.~Jones, {\em In-Plane Separations and High Momentum Structure 
            in d(e,e'p)n}, Jefferson Lab Proposal - E-94-004.
\bibitem{BF} S.~J.~Brodsky and G.~R.~Farrar, Phys.\ Rev.\ Lett.\  {\bf 31}, 1153 (1973).
\bibitem{Mueller}A.~H.~Mueller, Phys.\ Rept.\  {\bf 73}, 237 (1981).
\bibitem{FMS93}L.~Frankfurt, G.~A.~Miller and M.~Strikman, Nucl.\ Phys.\ A {\bf 555} 
               (1993) 752.
\bibitem{B82}S.~J.~Brodsky, in Proceedings if Thirteenth Intl. Symposium 
             on Multiparticle Dynamics, ed W.~Kittel, W.~Metzegar and A.~Stergiou, World 
             Scientific, Singapore,  1982, p.963.
\bibitem{M82}A.~H.~Mueller, in Proceedings if Seventeenth Rencontre de Moriond, 
             ed. J.~Tran Thanh Van, Editions Frontieres, Gif-sur-Yvette, France, 1982, 
             p.13.
\bibitem{JPR} P.~Jain, B.~Pire and J.~P.~Ralston, Phys.\ Rept.\  {\bf 271}, 67 (1996)
\bibitem{BFS}S.~J.~Brodsky, L.~Frankfurt, J.~F.~Gunion, A.~H.~Mueller and M.~Strikman,
             Phys.\ Rev.\ D {\bf 50}, 3134 (1994).
\bibitem{FMS93b}L.~Frankfurt, G.~A.~Miller and M.~Strikman, Phys.\ Lett.\ B {\bf 304}, 
                1 (1993)
\bibitem{Ashery}E.~M.~Aitala {\it et al.}  [E791 Collaboration], 
                Phys.\ Rev.\ Lett.\  {\bf 86}, 4773 (2001)

\bibitem{FGMSS96}L.~L.~Frankfurt, W.~R.~Greenberg, G.~A.~Miller, M.~M.~Sargsian and 
                M.~I.~Strikman, Phys.\ Lett.\ B {\bf 369}, 201 (1996).
\bibitem{FFLS}G.~R.~Farrar, H.~Liu, L.~L.~Frankfurt and M.~I.~Strikman, 
             Phys.\ Rev.\ Lett.\  {\bf 61} (1988) 686.
\bibitem{MJ} B.~K.~Jennings and G.~A.~Miller, Phys.\ Rev.\ Lett.\  {\bf 69},  3619 (1992)
\bibitem{FGMS}L.~Frankfurt, W.~R.~Greenberg, G.~A.~Miller and M.~Strikman,
              Phys.\ Rev.\ C {\bf 46}, 2547 (1992).
\bibitem{Nikol} N.~N.~Nikolaev, A.~Szczurek, J.~Speth, J.~Wambach, B.~G.~Zakharov and 
        V.~R.~Zoller, Nucl.\ Phys.\ A {\bf 567}, 781 (1994)
\bibitem{Kopel} B.~Z.~Kopeliovich, Sov.\ J.\ Nucl.\ Phys.\  {\bf 55}, 752 (1992).
\bibitem{BCT} M.~A.~Braun, C.~Ciofi degli Atti and D.~Treleani,
                      Phys.\ Rev.\ C {\bf 62}, 034606 (2000)
\bibitem{FMS92}L.~Frankfurt, G.~A.~Miller and M.~Strikman, 
               Comments Nucl.\ Part.\ Phys.\  {\bf 21}, 1 (1992).
\bibitem{chiral}L.~Frankfurt, T.S.H.~Lee, G.A.~Miller, M.~Strikman, Phys. Rev. 
                 {\bf C55}, 909-916 (1997).  
\bibitem{FPSS}L.~Frankfurt, E.~Piasetzky, M.~Sargsian, and M.~Strikman, 
              Phys. Rev. C {\bf 56}, 2752 (1997).
\bibitem{CFSM}C.~Ciofi degli Atti, L.~Frankfurt, S.~Simula and
              M.~Strikman, Phys. Rev. C {\bf 44}, 7 (1991). 
\bibitem{BBW}W. Bertozzi, W. Boeglin and L. Weinstein (spokespersons), 
            {\em Coincidence Reaction Studies with the LAS}, 
             Jefferson Lab Proposal - PR-89-015.
\bibitem{ESV}M.~Epstein, A.~Saha and E.~Voutier (spokespersons), 
             {\em Selected Studies of the 3He and 4He Nuclei through 
             Electrodisintegration at High Momentum Transfer}
	     Jefferson Lab Proposal - E-89-044.
\bibitem{Kim}E.~Egiyan (spokesperson), {\em Study of Short-Range Properties of 
             Nuclear Matter in Electron-Nucleus and Photon-Nucleus Interactions 
             with Backward Particle Production using the CLAS Detector},
             Jefferson Lab Proposal, PR-89-036.
\bibitem{Eli}W~Bertozzi, E.~Piasetzky and S.~Wood (spokespersons), 
             {\em Studying the Internal
             Small-Distance Structure of Nuclei via the Triple Coincidence (e,e'p+N) 
             Measurement}, Jefferson Lab Proposal - E-01-015.
\bibitem{Templon}J.~A.~Templon and J.~Mitchell (spokespersons), 
                {\em Systematic Probe of Short-Range Correlations}, 
                Jefferson  Lab Proposal - E-97-111.
\bibitem{vanLeeuwe} J.~J.~van Leeuwe {\it et al.}, Phys.\ Rev.\ Lett.\  {\bf 80}, 
                   2543 (1998).
\bibitem{Laget}J.M.~Laget, Nucl. Phys. A {\bf 579}, 333 (1994).
\bibitem{Rocco}R.~Schiavilla, Phys. Rev. Let. {\bf 65}, 835 (1990).
\bibitem{Nagorny}S.~I.~Nagorny {\em et al.}, Sov. J. Nucl. Phys. {\bf 53}, 228 (1991).
\bibitem{BNSUZ}O.~Benhar, N.~N.~Nikolaev, J.~Speth, A.~A.~Usmani and B.~G.~Zakharov,
               Nucl.\ Phys.\ A {\bf 673}, 241 (2000)
\bibitem{ABCKMT}M.~Alvioli, M.~A.~Braun, C.~Ciofi degli Atti, L.~P.~Kaptari, H.~Morita 
                and D.~Treleani, nucl-th/0012019, 2000.
\bibitem{Day}L.L.~Frankfurt, M.I.~Strikman, D.B.~Day and M.~Sargsian,
             Phys.\ Rev.\ C {\bf 48}, 2451 (1993).
\bibitem{FS91}L.~L.~Frankfurt, M.~I.~Strikman, {\em in Modern Topics in Electron 
              Scattering},Edited by B.~Frois and I.~Sick, World Scientific, 1991.
\bibitem{BJ}G.~E.~Brown and A.~D.~Jackson, {\em The Nucleon-Nucleon Interactions}, 
            North-Holland Publishing Company, 1976.
\bibitem{paris}M.~Lacombe, B.Loiseau, R. Vinh Mau, J.~Cote, P.~Pires and R. de Tourreil, 
              Phys. Lett. B {\bf 101}, 139 (1981).
\bibitem{bonn} R.Machleidt, K.~Holinde, and C.~Elster, Phys. Rep. {\bf 149}, 1 (1987).
\end{references}
\end{document}